\documentclass[aip,apl,twocolumn,nofootinbib,10pt]{revtex4-2}
\usepackage[T1]{fontenc}

\usepackage[dvipsnames]{xcolor}
\usepackage{color}

\definecolor{PPQAblue}{HTML}{045275} 
\definecolor{PPQAgreen}{HTML}{089099}
\definecolor{PPQAlightgreen}{HTML}{7CCBA2}
\definecolor{PPQAorange}{HTML}{F0746E}
\definecolor{PPQApink}{HTML}{DC3977} 
\definecolor{PPQApurple}{HTML}{7C1D6F}

\usepackage[normalem]{ulem}
\usepackage{pgfplots}
\pgfplotsset{compat = newest}
\usepackage{adjustbox}
\usepackage[utf8]{inputenc}
\usepackage{physics}
\usepackage{bm} 
\usepackage{bbm} 
\usepackage{braket}
\usepackage{dsfont}
\usepackage{amsmath}
\usepackage{amssymb}
\usepackage{amsthm}
\usepackage{thmtools} 
\usepackage{mathtools}
\usepackage[switch]{lineno} 
\usepackage[ruled,linesnumbered]{algorithm2e}
\SetArgSty{textnormal}
\usepackage{algpseudocode}
\usepackage{subfiles}
\usepackage{tikz}
\usetikzlibrary{quantikz}
\usetikzlibrary{shapes.geometric,backgrounds,positioning,shapes.geometric,decorations.markings,decorations.pathreplacing,arrows,knots,hobby,angles,quotes,trees}
\tikzset{meter/.append style={draw, inner sep=5, rectangle, font=\vphantom{A}, minimum width=20, line width=.4,
 path picture={\draw[black] ([shift={(.1,.2)}]path picture bounding box.south west) to[bend left=50] ([shift={(-.1,.2)}]path picture bounding box.south east);\draw[black,-latex] ([shift={(0,.1)}]path picture bounding box.south) -- ([shift={(.3,-.1)}]path picture bounding box.north);}}} 
\makeatletter
\newcommand{\gettikzxy}[3]{%
  \tikz@scan@one@point\pgfutil@firstofone#1\relax
  \edef#2{\the\pgf@x}%
  \edef#3{\the\pgf@y}%
}
\makeatother
\usepackage{graphics}
\usepackage{graphicx}
\usepackage{float}
\usepackage{tcolorbox}  
\usepackage{ragged2e}
\usepackage{enumitem}
\usepackage[colorlinks]{hyperref}
\hypersetup{
    colorlinks  = true,
    citecolor   = PPQAblue,
    linkcolor   = PPQAgreen,
    urlcolor    = PPQAblue
}

\apptocmd{\thebibliography}{\small}{}{}

\setcitestyle{numbers,square}

\usepackage{silence}
\usepackage{etoolbox}
\apptocmd{\sloppy}{\hbadness 10000\relax}{}{}

        \DeclareMathOperator{\GHz}{GHz} 
        \DeclareMathOperator{\GB}{GB} 
        \DeclareMathOperator{\second}{s} 

        \DeclareMathOperator{\KP}{KP} 
        \DeclareMathOperator*{\opt}{opt} 
        \DeclareMathOperator*{\out}{out} 
        \DeclareMathOperator*{\total}{tot} 
        \DeclareMathOperator*{\ancilla}{a} 
        
        \newcommand*{\N}{\mathbb{N}}

        \newcommand*{\C}{\mathbb{C}}
        \newcommand*{\hil}{\mathcal{H}}
        \newcommand*{\defcolon}{\,:\,} 
        \newcommand*{\hamming}{\Delta} 
        \newcommand*{\numbits}[1]{\abs{{#1}_{\text{bin}}}} 
        \newcommand*{\scriptin}{\raisebox{0.15ex}{$\scriptscriptstyle\in$}} 
        
        \newcommand*{\kp}{\KP_{n}(\bm{p}, \bm{w}; c)} 
        \newcommand*{\kpex}{\KP_{4}(\bm{p}, \bm{w}; 7)} 
        \newcommand*{\QTG}{\mathcal{G}} 
        \newcommand*{\controlled}{C} 
        \newcommand*{\QSop}{\mathcal{Q}} 
        \newcommand*{\signflip}{\mathcal{S}} 
        \newcommand*{\qubitc}{\ensuremath{\mathbf{Q}}} 
        \newcommand*{\gatec}{\ensuremath{\mathbf{G}}} 
        \newcommand*{\cyclec}{\ensuremath{\mathbf{C}}} 
\newcommand*{\SW}[1]{{\color{purple}[SW: #1]}}

\newcommand*{\HL}[1]{#1}

\AtBeginDocument{\renewcommand{\arccos}{\operatorname{acos}}}
\selectfont

\begin{document}

\begin{abstract}
    Here we present two novel contributions for achieving quantum advantage in solving difficult optimisation problems, both in theory and foreseeable practice.
    (1) We introduce the ``Quantum Tree Generator'', an approach to generate in superposition all feasible solutions of a given instance, yielding together with amplitude amplification the optimal solutions for $0$-$1$ knapsack problems.
    \HL{The QTG offers massive memory savings and enables competitive runtimes compared to the classical state-of-the-art knapsack solvers (such as COMBO, Gurobi, CP-SAT, Greedy) already for instances involving as few as 100 variables.
    (2) By introducing a new runtime calculation technique that exploits logging data from the classical solver COMBO, we can predict the runtime of our method way beyond the range of existing quantum platforms and simulators, for various benchmark instances with up to 600 variables.}
    Combining both of these innovations, we demonstrate the QTG's potential \HL{\emph{practical} quantum} advantage for large-scale problems, indicating an effective approach for combinatorial optimisation problems. 
\end{abstract}

\title{A quantum algorithm for solving 0-1 Knapsack problems}

\author{S\"oren Wilkening}\email{soeren.wilkening@itp.uni-hannover.de}
\affiliation{Institut f\"ur Theoretische Physik, Leibniz Universit\"at Hannover, Germany}
\affiliation{Volkswagen AG, Berliner Ring 2, 38440 Wolfsburg}
\author{Andreea-Iulia Lefterovici}
\affiliation{Institut f\"ur Theoretische Physik, Leibniz Universit\"at Hannover, Germany}
\author{Lennart Binkowski}
\affiliation{Institut f\"ur Theoretische Physik, Leibniz Universit\"at Hannover, Germany}
\author{Michael Perk}
\affiliation{Institut f\"ur Betriebssysteme und Rechnerverbund, Technische Universität Braunschweig, Germany}
\author{Sándor P. Fekete}
\affiliation{Institut f\"ur Betriebssysteme und Rechnerverbund, Technische Universität Braunschweig, Germany}
\author{Tobias J. Osborne}
\affiliation{Institut f\"ur Theoretische Physik, Leibniz Universit\"at Hannover, Germany}

\maketitle

\section{\label{section:Introduction}Introduction}
\begin{figure*}[!ht]
    \begin{tcolorbox}[
        colback=white, 
        colframe=PPQAgreen!25,
        coltitle=black, 
        title=\noindent\justifying{Box 1: Quantum Tree Generator}]
    \begin{minipage}[t]{0.47\textwidth}
        \justifying\footnotesize
        \noindent\textbf{Input:}
        
        \noindent Knapsack instance $\kp$, feasible assignment $\bm{x}$
        \vspace*{8pt}
        
        \noindent\textbf{1. Initialize:}
        
        \noindent Initialize the three registers in the state $\ket{\bm{0}}^{1} \ket{c}^{2} \ket{0}^{3}$.
        \vspace*{8pt}
        
        \noindent\textbf{2. Tree traversing:}
        
        \noindent For each item $m = 1, \ldots, n$, perform the following steps:
        \vspace*{5pt}
        
        \noindent\textbf{(a)} Create a superposition. Apply biased Hadamard H$^\prime$ on the $m$-th qubit in register 1, controlled on whether $w_{m}$ does not exceed the remaining capacity stored in the register 2.
        \vspace*{5pt}
        
        \noindent\textbf{(b)} Update the remaining capacities. Subtract $w_{m}$ from the state of register 2, controlled on the $m$-th qubit in register 1.
        \vspace*{5pt}
        
        \noindent\textbf{(c)} Update the total profit. Add $p_{m}$ to the state of register 3, controlled on the $m$-th qubits in register 1.
        
    \end{minipage}\hfill
    \begin{minipage}[t]{0.5\textwidth}
        \justifying\footnotesize
        \noindent\textbf{Example:}
        
        \noindent Knapsack instance $\kpex$ with weights $\bm{w}$ = (2, 2, 1, 5) and profits $\bm{p}$ = (6, 2, 1, 2). 
        H' is a biased Hadamard gate.
        \begin{center}
            \includegraphics[width=\textwidth]{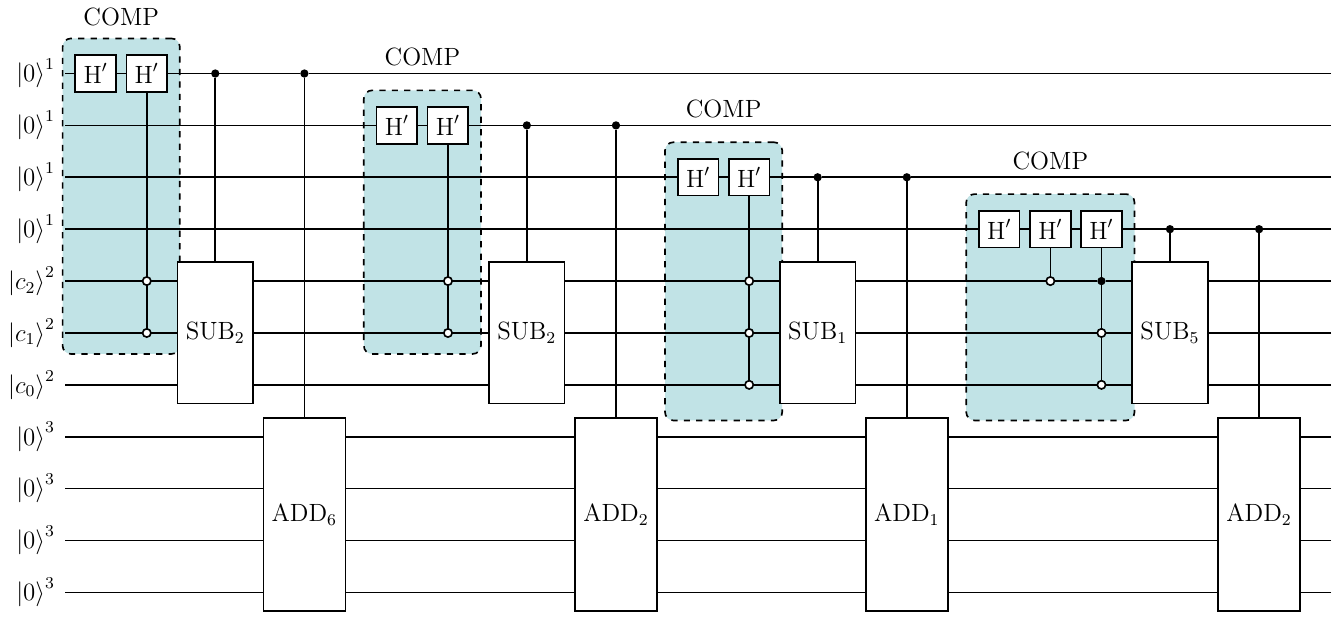}
        \end{center}
    \end{minipage}
    \label{box:QTG}
    \end{tcolorbox}
\end{figure*}
\vspace*{-3mm}

\HL{In recent years, the prospect of real-world quantum computing has raised hopes for solving hard combinatorial optimisation problems, leading to tremendous theoretical work on developing and analysing suitable quantum methods.
On the one hand, it is reasonable to expect in the next years that quantum information processing devices will comprise many thousands of physical qubits~\cite{ResearchIBM2023Charting100000Qubits}.
On the other hand, this progress is inhibited by two major challenges.} 

\HL{At the lower level, one challenge emerges from the effects of decoherence \cite{Brandt1999QubitDevicesAndTheIssueOfQuantumDecoherence, Schlosshauer2019QuantumDecoherence}.
The current error rates experienced by all extant quantum devices place them well above the fault tolerance threshold \cite{Preskill2018QuantumComputingInTheNISQEraAndBeyond} required for the effective deployment of quantum error correction \cite{Lidar2013QuantumErrorCorrection}.
Further critical roadblocks, depending on platform \cite{Kjaergaard2020AuperconductingQubits,Siddiqi2021EngineeringHighCoherenceSuperconductingQubits, Blatt2012QuantumsimulationsWithTrappedIons, Cirac1995QuantumComputationWithColdTrappedIons, Haffner2008QuantumComputingWithTrappedIons, Leibfried2003QuantumDynamicsofSingleTrappedIons, Madsen2022QuantumComputationalAdvantageaWithAProgrammablePhotnicProcessor,Slussarenko2019PhotonicQuantumInformationProcessingAConciseReview, Wu2020QuantumComputingwithMultidimensionalClusterStatesInPhotonicPlatform}, also include the difficulties of implementing mid-circuit measurements \cite{Rudinger2022CharacterizingMidcircuitMeasurementsOnASuperconductingQubitUsingGateSetTomography} and entangling gates \cite{Reagor2018DemonstrationOfUniveralParametricEntanglingGateOnAMultiQubitLattice}.
That quantum computers deliver useful improved and accelerated solutions is expected in the long term when logical qubits are plentiful and cheap.
However, in the near- and mid-term, the situation is subtle and nuanced. Justifying the considerable effort for ensuing research and development hinges on the expected practical usefulness of quantum devices.}

\HL{At a higher level, the challenge lies in demonstrating quantum advantage in practical applications.
In recent times, there has indeed been considerable optimism that combinatorial optimisation problems provide a rich class of application areas whose solutions quantum computers will improve~\cite{Ajagekar2020QuantumComputingBasedHybridSolutionStrategiesForLargeScaleDiscreteContinuousOptimizationProblems}. 
Although quantum devices have already demonstrated accelerated solutions for synthetic problems \cite{Arute2019QuantumSupremacyUsingAProgrammableSuperconductingProcessor}, there has been, to date, no conclusive demonstration of a practical quantum advantage for realistic instances in comparison with the \emph{best classical approach}.
Quantum heuristics \cite{VanDam2021QuantumOptimizationHeuristicsWithAnApplicationToKnapsackProblems} currently do not provide competitive results when compared with classical solvers, such as CPLEX~\cite{CPLEX}, GUROBI~\cite{Gurobi} and CP-SAT~\cite{CPSAT} which routinely solve problems involving thousands of variables to \emph{provable optimality}.}

\HL{On the quantum side, the development of new algorithms has been almost exclusively driven by theoretical measures, in particular by worst-case, asymptotic runtime analysis. 
In particular, Grover's algorithm~\cite{Grover1996AFastQuantumMechanicalAlgorithmForDatabaseSearch} and its provable quadratic speedup over unstructured linear search have spawned a rich manifold of quantum routines
\cite{Brassard2002QuantumAmplitudeAmplificationAndEstimation,Cerf2000NestedQuantumSearchAndStructuredProbleems,Montanaro2018QuantumWalkSpeedupOfBacktrackingAlgorithms, Montanaro2020QuantumSpeedupOfBranchAndBoundAlgorithms}.
However, these theoretical insights have to be taken with a grain of salt when it comes to their practical applicability:
A favourable asymptotic worst-case complexity does not necessarily imply a speedup for instance sizes that are relevant in practice.
This already holds true for the comparison of two purely classical algorithms.
The simplex method~\cite{Danzig1948LinearProgrammingInProblemsForTheNumericalAnalysisOfTheFuture} for solving linear programs has a worst-case runtime exponential in the input dimension, but remains the best-performing algorithm in practice.
In contrast, the ellipsoid method~\cite{Khachiyan1979APolynomialAlgorithmInLinearProgramming} with worst-case runtime polynomial in the input dimension, performs poorly in practice in most cases, especially when compared to the simplex method~\cite{Yinyu1997InteriorPointAlgorithms}.
This immanent gap between asymptotic worst-case behaviour and practical performance does not necessarily shrink when the compared algorithms run on completely different architectures.}

\HL{Thus, we are faced with a fundamental dilemma when exploring the possible real-world impact of quantum methods for practical optimisation: On the one hand, the justification for building and tuning quantum devices hinges on their practical impact; on the other hand, gauging their usefulness by running real-world benchmarks is impossible before these devices exist. A possible recourse for the latter is to perform classical simulations of quantum methods; however, possibilities for general gate-based simulations are saturated both in memory and runtime at around $50$~qubits~\cite{Lykov2023FastSimulationOfHighDepthQAOACircuits}, which is far too few to evaluate the performance of a quantum heuristic against realistic benchmark instances (which typically would require between $100$-$10000$ qubits).}

\HL{One recent approach \cite{Cade2022QuantifyingGroverSpeedupsBeyondAsymptoticAnalisys} for obtaining realistic estimates for runtimes of quantum algorithms beyond asymptotic scaling is to derive quantitative complexity bounds for quantum routines, rather than asymptotic worst-case $\mathcal{O}$-expressions.
These bounds are fed with data obtained by running classical versions of these routines.
In this light, a recent study \cite{Ammann2023RealisticRuntimeAnalysisForQuantumSimplexComputation} was able to show that the quantum version of the simplex method \cite{Nannicini2024FastQuantumSubroutinesSimplex}, despite its promising asymptotic runtime, cannot outperform the standard simplex method on any reasonable benchmark instance.
While this framework is powerful for benchmarking quantum routines when having formulas for the expected runtime, it does not provide enough tools for analysing quantum heuristics.}

\HL{In this paper, we directly take on the twin challenges of (1) developing a quantum heuristic method to solve combinatorial optimisation problems and (2) evaluating the performance of this heuristic method against benchmark instances.
We target the knapsack problem -- the hydrogen atom of combinatorial optimisation problems -- which frequently appears in real-world applications such as portfolio optimisation and securitisation \cite{Kellerer2004KnapsackProblems}, and often arises as a subproblem in more complex problems such as resource allocation and scheduling.
For this fundamental problem, we develop a quantum method we name Quantum Tree Generator (QTG) to generate all feasible solutions in superposition, which yields the optimal solution in concert with quantum amplitude amplification~\cite{Brassard2002QuantumAmplitudeAmplificationAndEstimation}.
In order to evaluate the performance of the QTG-based search on realistic benchmark instances, we also introduce a novel technique to compute the quantum algorithm's runtime:
Extracting crucial logging data from the classical solution, obtained via the current champion COMBO \cite{Martello1999DynamicProgrammingAndStrongBoundsForTheKnapsackProblem} method, we can infer the expected number of cycles required by the QTG-based method to solve the same instance (importantly, we cannot simulate the method, only infer the expected runtime in elementary gates using this side information).
Our benchmarking leads us to predict that the QTG-based method requires fewer cycles to solve realistic instances already at $100$ variables.
Most dramatic, however, are the memory requirements:
The COMBO method -- based in part on dynamic programming -- can make huge demands on memory, with $10^{10}$ bits being routinely requested.
The QTG-based search, however, requires only a constant multiple of the variable number in logical qubits.
This leads to a possible quantum advantage in both time and space starting at as few as $100$ variables, offering a clear perspective of a \emph{practical} quantum advantage for instance sizes of real-world relevance.}


\section{\label{section:PreviousWork}Previous Work}
\HL{In 1995, Pisinger developed the branch-and-bound approach called EXPKNAP \cite{Pisinger1995AnExpandingCoreAlgorithmForTheExactKnapsackProblem}.
At the time, the algorithm stood out as one of the most efficient algorithms documented in literature for the 0-1 knapsack problem (0-1-$\KP$).
In recent years, integer programming (IP) and constraint programming (CP) solvers emerged as the industry standard for various optimisation problems. The integer programming solver GUROBI (based on branch and cut methods) and CP-SAT \cite{CPSAT} (portfolio solver based on constraint programming) are popular solvers in their respective field.
In 1999, the COMBO algorithm emerged as a dynamic programming solver specifically designed for 0-1-$\KP$.
Known for its stable behaviour due to its pseudo-polynomial time complexity bound  \cite{Martello1999DynamicProgrammingAndStrongBoundsForTheKnapsackProblem}, it stands as the current leading solver for this problem.}

\begin{figure*}[!ht]
    \centering
    \begin{minipage}{\textwidth}
        \centering
        \includegraphics[width=\textwidth]{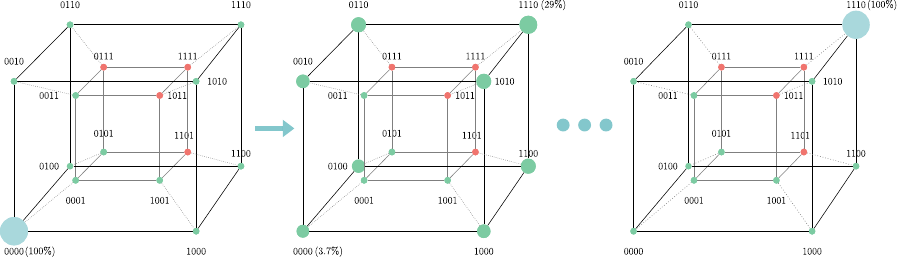}
    \end{minipage}\hfill
    \caption{\label{figure:Hypercube}
        The action of the QTG on the $\kpex$ example.
        The path register holds 16 computational basis states -- here depicted as the corners of a hypercube -- representing all feasible (green dots) and infeasible (orange dots) four-item paths.
        The state $\ket{1110}$ represents the optimal solution.
        The initial state of the system is $\ket{0000}$, corresponding to an entirely empty knapsack.
        After one application of the QTG, the optimal state's sampling probability is increased to $29\%$.
        After 12 applications of the QTG, the optimal state is reached.
    }
\end{figure*}

\HL{Quantum methods to tackle constrained optimisation (and closely related satisfaction) problems have been previously studied in works such as nested quantum search \cite{Cerf2000NestedQuantumSearchAndStructuredProbleems} and quantum branch-and-bound  \cite{Montanaro2020QuantumSpeedupOfBranchAndBoundAlgorithms, Chakrabarti2022UniversalQuantumSpeedupForBranchAndBoundBranchAndCutAndTreeSearchAlgorithms}.
Nested quantum search is an enhancement over Grover's \cite{Grover1996AFastQuantumMechanicalAlgorithmForDatabaseSearch} unstructured search algorithm for solving constrained satisfaction problems.
It consists of applying Grover's search on a sub-problem to identify potential solutions, followed by another Grover's search application on the remaining variables to extract the actual solutions from the set of potential solutions.
This prepares a uniform superposition over the set of feasible assignments.
The number of iterations for the nested quantum search to find the optimal solution is $\mathcal{O}(\sqrt{d^\alpha})$, where $d$ is the dimension of the search space, and $\alpha < 1$ is a constant depending on the nesting depth.
Quantum branch-and-bound (QBnB) uses quantum backtracking  \cite{Montanaro2018QuantumWalkSpeedupOfBacktrackingAlgorithms, Martiel2020PracticalImplementationQuantumBacktrackingAlgorithm} as a subroutine.
The algorithm improves upon nested quantum search by searching at each level of the tree only for partial assignments that satisfy the constraints and show promise to deliver optimal solutions based on unassigned variables.
The number of iterations for the quantum branch-and-bound methods to find the optimal solution is $\mathcal{O}(\sqrt{Tp})$, where $T$ is the number of nodes and $p$ is the depth of the tree.
Both algorithms achieve a quadratic speedup over their respective classical counter parts.
We benchmark these methods in \autoref{Section:BenchmarkingOfCompetitiveMethods}.}

\HL{In contrast, our method prepares a superposition of all feasible states by sequentially imposing the constraints of an instance (as shown in \autoref{figure:Hypercube}), 
followed by an augmented version of amplitude amplification to get the optimal solution.}

\section{\label{section:Preliminaries}Preliminaries}
Consider a set of $n$ items, each of which has some integer profit $p_{m} > 0$ and \HL{weight $w_{m} > 0$.}
The objective is to select a subset of items of maximum cumulative profit, so that their cumulative weight do not exceed some threshold \HL{$c > 0$).
An instance of the 0-1-KP is completely defined by the number $n$ of items, a list $\bm{p} \coloneqq (p_{1}, \ldots, p_{n})$ of profit values, \HL{a list $\bm{w} \coloneqq (w_{1}, \ldots, w_{n})$ of weights, and the capacity $c$};
we address a given instance in the following by $\kp$.
It has the following formulation as an Integer Linear Program (ILP):}
\begin{linenomath}
\HL{\begin{align}\label{equation:KnapsackProblem}
\begin{split}
    \text{maximise } & \sum_{m = 1}^{n} p_{m} x_{m} \\
    \text{subject to } & \sum_{m = 1}^{n} w_{m} x_{m} \leq c \\
    x_{m} \in\{0, 1\}, & \quad m = 1, \ldots, n.
\end{split}
\end{align}}
\end{linenomath}
\HL{For the 0-1-KP, \HL{the binary variable $x_{m}$} encodes the choice of either including item $m$ into the knapsack ($x_{m} = 1$) or omitting it ($x_{m} = 0$).}
Therefore, a complete assignment of all items (also called \emph{path}) constitutes a bit string of length $n$.

The standard procedure to formulate the 0-1-KP for study via a quantum computer is to assign one qubit to each decision variable \HL{$x_{m}$, thus representing the paths $\bm{x} = x_{1} \ldots x_{n}$ as computational basis states $\ket{\bm{x}}^{1}$} in an $n$-qubit \emph{path register} $\hil_{1} = \C^{2^{n}}$.
Additionally, for any feasible assignment \HL{$\bm{x}$, we store the binary representation of the remaining capacity $c - \sum_{m = 1}^{n} w_{m} x_{m}$ as a quantum state $\ket{c_{\bm{x}}}^{2}$ in a $\numbits{c}$-qubit \emph{capacity register} $\hil_{2}$}.\footnote{With $\numbits{a} = \lfloor\log_{2} a\rfloor + 1$ we denote the length of the binary representation of $a$.} 
\HL{The total profit is stored as a state $\ket{P_{\bm{x}}}^{3}$ in a $\numbits{P}$-qubit \emph{profit register} $\hil_{3}$}, where $P$ is any upper bound on the optimal profit.
\HL{Furthermore, we optimise the depth of the circuit by using a $\max(n, \numbits{c}, \numbits{P})$-qubit ancilla register $\hil_{\ancilla}$.
The total number of used qubits in the composite register $\hil = \hil_{1} \otimes \hil_{2} \otimes \hil_{3} \otimes \hil_{\ancilla}$ is therefore given by
\begin{equation}\label{equation:TotalNumberOfQubits}
    n + \numbits{c} + \numbits{P} + \max(n, \numbits{c}, \numbits{P}).
\end{equation}}

Our quantum algorithm for solving 0-1-KP operates entirely on the aforementioned composite register.
The quantum routine is fundamentally based on two main ingredients:
\begin{itemize}
    \item[1.] a state preparation circuit that creates a superposition of all valid paths along with their total profit and remaining capacity as composite quantum states.
    \item[2.] an augmented quantum amplitude amplification \cite{Brassard2002QuantumAmplitudeAmplificationAndEstimation} which increases the success probability of measuring the optimal solution while avoiding endless loops.
\end{itemize}


\section{\label{section:Method}Methods}
\begin{figure*}
    \includegraphics[width=\linewidth]{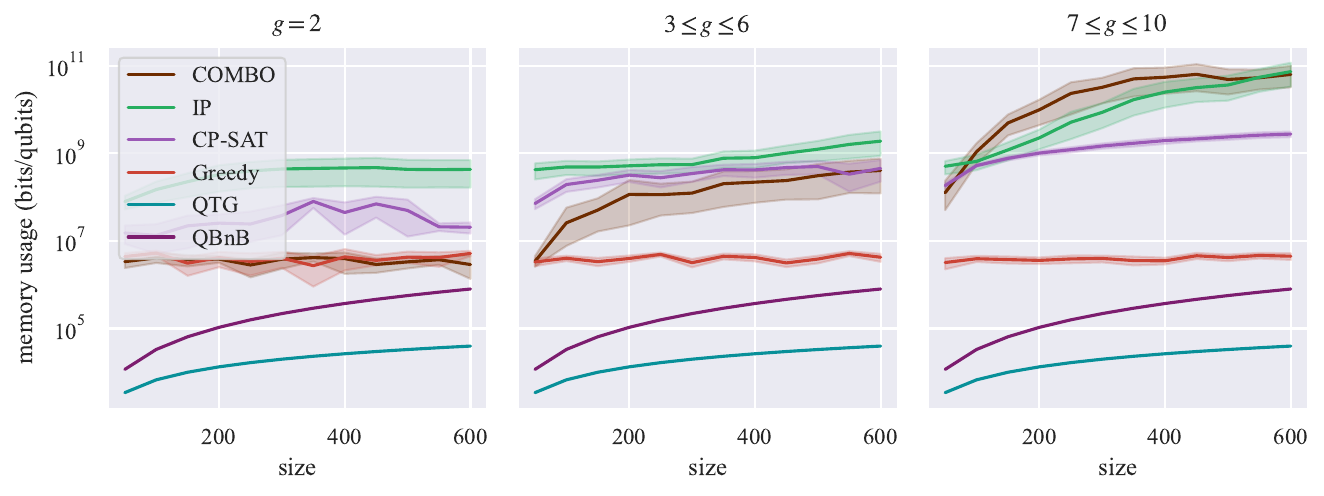}
    \caption{\HL{Memory requirements for quantum methods (QBnB and QTG-based search) in comparison with standard classical solvers. IP (Gurobi) and CP-SAT were executed for the same time as it took COMBO to find OPT to have a fairer comparison of the used memory. As a result, these solvers were unable to find the optimal solution for all benchmark instances.}}
    \label{figure:QTGMemoryBenchmark}
\end{figure*}

The entire state preparation circuit, the Quantum Tree Generator (QTG) $\QTG$, arises as a cascade of $n$ \emph{layer unitaries} $U_{m}$, i.e., $\QTG = \prod_{m = 1}^{n} U_{m}$.
Applying the QTG to the initial state \HL{$\ket{\bm{0}}^{1} \ket{c}^{2} \ket{0}^{3}$} generates a (generally non-uniform) superposition of all valid paths.
The QTG iteratively includes/excludes all items in a weighted superposition while checking for feasibility and updating the capacity and profit registers.
First, for every path, it is checked whether the remaining capacity covers the weight of the current item $m$. 
Next, every path that can include the item will be branched using a biased Hadamard gate H$^{x_{m}}_{b}$
\footnote{For brevity, we also write H' when the parameters are either clear from the context or not relevant.}
defined as
\begin{linenomath}
\HL{\begin{equation}\label{equation:biasedHadamard}
\begin{split}
     \text{H}^{0}_{b}\hspace*{-0.1cm} &=\hspace*{-0.1cm} \text{Ry}\bigg(2 \arccos{\sqrt{\frac{1 + b}{2 + b}}}\bigg)\hspace*{-0.1cm} =\hspace*{-0.1cm} \frac{1}{\sqrt{2 + b}} 
    \begin{pmatrix}
        \sqrt{1 + b} & -1 \\
        1 & \sqrt{1 + b}
    \end{pmatrix},\\
    \text{H}^{1}_{b}\hspace*{-0.1cm} &=\hspace*{-0.1cm} \text{Ry}\bigg(2 \arccos{\sqrt{\frac{1}{2 + b}}}\bigg)\hspace*{-0.1cm} =\hspace*{-0.1cm} \frac{1}{\sqrt{2 + b}} \begin{pmatrix}
        1 & -\sqrt{1 + b} \\
        \sqrt{1 + b} & 1
    \end{pmatrix},
\end{split}
\end{equation}}
\end{linenomath}
\HL{where $x_{m}$ is the $m$-th bit of an intermediate solution $\bm{x}$.}
Upon successful item inclusion, the item's weight is subtracted from the capacity register's state and its profit is added to the profit register's state.
Both, subtraction and addition, are conducted using Quantum Fourier Transform (QFT) adders \cite{Draper2000AdditionOnAQuantumComputer}.
In total, the unitaries $U_{m}$ are comprised of three components \HL{$U^{\vphantom{1}}_{m} = U^{3}_{m}\, U^{2}_{m}\, U^{1}_{m}$:}
\begin{linenomath}
\HL{\begin{align}
\begin{split}
    U^{1}_{m} &= \controlled^{2}_{\geq w_{m}}  \left(\text{H}^{x_{m}}_{b}\right) \\
    U^{2}_{m} &= \controlled^{1}_{m}\left( \text{SUB}_{w_{m}} \right)\\
    U^{3}_{m} &= \controlled^{1}_{m}\left( \text{ADD}_{p_{m}} \right).
\end{split}
\end{align}}
\end{linenomath}
\HL{We denote the control of the biased Hadamard gates with $C_{\geq w_{m}}^{2}$ because it asks whether the second register's state is greater or equal to the integer $w_{m}$.
Subtract and addition are controlled on the $m$-th qubit in the first register: $\controlled^{1}_{m}$.
The bias $b$ in order to maximise the sampling probability of a path $\bm{x}$ within the superposition created by the QTG is given by
\begin{linenomath}
\begin{align}\label{equation:optimalBias}
    b_{\opt} = \frac{n}{\hamming} - 2 \approx \frac{n}{\hamming},
\end{align}
\end{linenomath}
where $\hamming \coloneqq \hamming(\bm{x}',\bm{x})$ is the Hamming distance of $\bm{x}$ to a given intermediate solution $\bm{x}'$.}
More details on the construction of the biased Hadamard gates H$^{x_{m}}_{b}$ and unitaries $U_{m}$ can be found in \autoref{subsection:BranchingStrategy}.
The concept of the QTG is summarised in \hyperref[box:QTG]{Box 1}, along with the $\kpex$ circuit example.

\begin{figure*}
    \includegraphics[width=\linewidth]{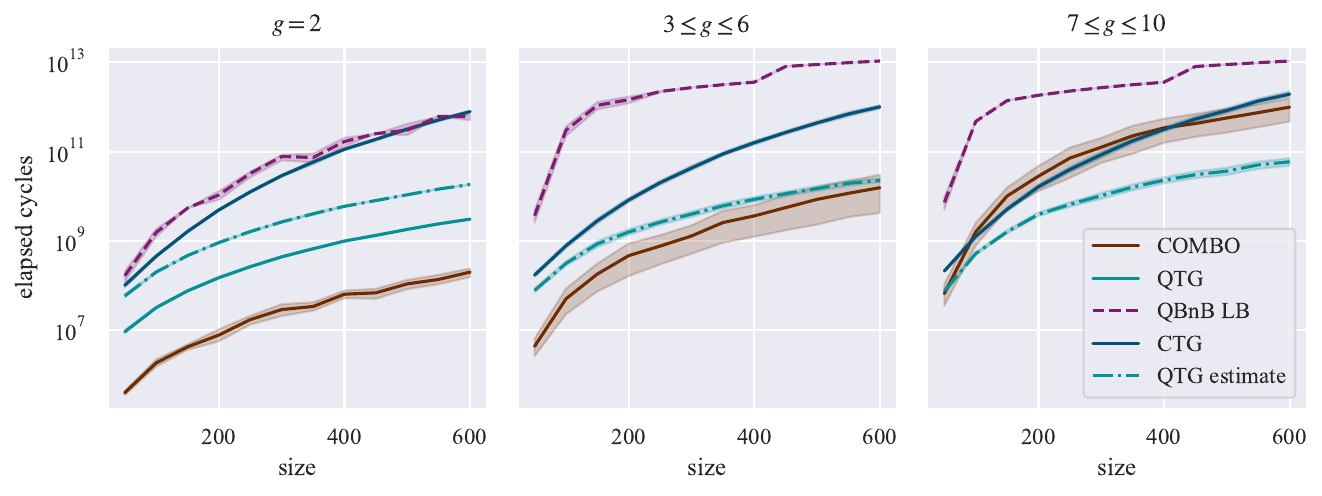}
    \caption{\HL{Comparison of quantum methods (QBnB and QTG-based search) with the classical methods COMBO and CTG, for different group numbers $g$. Note that the values for QTG are realistic estimates, based on simulation (for $g = 2$) and calibrated estimates (for $g \geq 2$); on the other hand, values for QBnB are lower bounds, based on several benevolent assumptions, so a realistic runtime can be expected to be higher; see~\autoref{Section:Qbnb}. For larger values of $g$, computational effort for full QTG simulation is excessive, so we provide estimates for QTG derived from CTG and justified in \autoref{subsubsection:BiasedHadamardGates}.}}
    \label{figure:QTGRuntimeBenchmark}
\end{figure*}

The superposition created by the QTG is further manipulated in order to increase the probability of measuring the optimal solution states.
For this, we employ quantum maximum finding (\textbf{QMaxSearch}) \cite{Durr1996AQuantumAlgorithmForFindingTheMinimum}, which iteratively calls an amplitude amplification (\textbf{QSearch}) \cite{Brassard2002QuantumAmplitudeAmplificationAndEstimation} routine.
The intermediate best solution's total profit provides the current threshold and is therefore entirely evaluated on the third register.
More details can be found in \autoref{subsection:QuantumMaximumFinding}.

\HL{We further choose a bias of $b = n / 4$, based on the calculations in \autoref{subsection:BranchingStrategy}, and a maximum number of $M = 700 + n^{2} / 16$ Grover iterations within \textbf{QSearch}.
In a classical preprocessing step, we sort the list of items by decreasing density (profit versus cost), and obtain an initial threshold for \textbf{QMaxSearch} by greedily packing as many items into the knapsack as possible (this approach is also called Greedy).}

By combining \HL{the application} of the QTG (see \autoref{figure:Hypercube} for the $\kpex$ example), the phase flip operator within \textbf{QSearch}, and quantum measurements with a detailed analysis of the resource requirements of every subcircuit, we calculate the number of necessary quantum cycles for obtaining the optimal solution with a certain success probability.
More details can be found in \autoref{subsection:RuntimeCalculator} and \autoref{subsection:ResourceEstimation}.
Our analysis makes the following crucial assumptions:
\begin{itemize}
    \item[1.] all qubits and gates are noiseless, logical components.
    \item[2.] all single-qubit gates, single-controlled single-qubit rotations, and Toffoli gates together constitute a universal set of elementary gates of equal implementation cost.
    \item[3.] disjoint gates can be executed in parallel, constituting a single cycle on the QPU.
\end{itemize}

\HL{Lastly, the QTG can be dequantised in order to yield a probabilistic classical algorithm:
Starting from the all-zero string $\bm{0}$, iterate through all items/bits $x_{m}$ and apply conditioned bit flips with probability $\abs{(\text{H}_{b}^{x_{m}})_{1 2}}^{2}$.
The condition $C_{\geq w_{m}}^{2}$ can be realised classically.
The entire algorithm then consists of sampling multiple times from the generated distribution of bit strings and updating the current solution whenever a better candidate was sampled. 
More details can be found in \autoref{subsubsection:BiasedHadamardGates}.
We name this quantum-inspired method the \emph{Classical Tree Generator} (CTG).
Similar (although not quantum-inspired) constructions already exist in the literature~\cite{Jooken2023MonteCarloTreeSearchAndInstanceSpaceAnalysisForTheKnapsackProblem}.}


\section{\label{section:Results}Results}
\HL{We demonstrate the capabilities of our QTG-based search on a diverse set of benchmark instances.
All experiments were carried out on a regular desktop workstation with an
AMD Ryzen 7 5800X ($8\times 3.8 \GHz$) CPU and $128 \GB$ of RAM\@.
Instances were produced using a state-of-the-art instance generator from \citet{Jooken2022ANewClassOfHardProblemInstancesForThe0_1KnapsackProblem} that was made available in~\cite{Jori2020KnapsackProblemInstances}.
The problem instances involve items partitioned into groups with exponentially decreasing profits and weights.
Items within the same group have similar, slightly perturbed profits and weights. The last group differs by having uncorrelated small profits and weights, introducing a variety of profit-weight ratios.
The generator created six instances for every value $2 \leq g \leq 10$ (number of groups) and size $n = 50, 100, \ldots, 600$. All instances were generated with capacity $c = 10\,000\,000\,000$. This yielded a total of $648$ benchmark instances with various difficulties.}

\HL{We compare COMBO's and CTG's classical runtime as well as runtime calculations for the QTG-bases search and the quantum branch-and-bound method proposed by \citet{Chakrabarti2022UniversalQuantumSpeedupForBranchAndBoundBranchAndCutAndTreeSearchAlgorithms}.
To justify the use of COMBO, in \autoref{subsubsection:ClassicalMethods} we compare the capabilities of different classical knapsack solvers.
The code and data are publicly available\footnote{Source code and data: \url{https://github.com/SoerenWilkening/QTG_0-1Knapsack}}.}

\HL{COMBO is a single-core algorithm, making comparing elapsed cycles and memory usage a fair unit of measurement.
To measure the elapsed cycles of COMBO and CTG, we read the time stamp counter (TSC) values, that are present on x86 processors.
The required memory (resident set size) was measured with GNU time utility.
In contrast to quantum systems, classical computers use a certain amount memory for storing the code of the program.
To calculate the data memory usage, all solvers were executed on tiny instances ($10$ items). 
This makes sure that all relevant code memory pages were loaded by the solvers at some point and therefore detected by the GNU time utility.
The data memory usage is now calculated by subtracting the minimum memory usage of that solver in the preliminary experiment.}

\HL{Because QTG-based search is a non-deterministic algorithm, the optimal solution is only sampled with a certain success probability recorded for every instance.
For the easier instances with $g = 2$, we employ both the exact simulation for QTG, as well as a hybrid version (QTG estimate), for which we execute the CTG with the square of the prescribed number of Grover iterations to determine the success probability.
For harder instances, we omit the exact simulation due to long processing times, but keep the QTG estimate (see~\autoref{subsubsection:BiasedHadamardGates}), which shows the same scaling behaviour as the exact calculator but with a positive offset, making it a heuristic upper bound. All this provides realistic evaluations for QTG, based on exact simulations (for easier instances up to 600 items) and calibrated pessimistic estimates (for harder instances up to 600 items).}

\HL{This differs from the runtime estimates for QBnB, which are generous lower bounds for any practical runtime, based on several benevolent assumptions; a realistic runtime of QBnB can be expected to be considerably higher. These assumptions for QBnB (which are \emph{not} granted to QTG) include a success probability of $1$ for all quantum subroutines; see \autoref{Section:Qbnb} for details. As the size of the tree grows exponentially, we limit the maximum number of tree nodes to $10^{10}$ giving us an even more optimistic lower bound on the actual number of cycles.}

\begin{figure}
    \includegraphics[width=\linewidth]{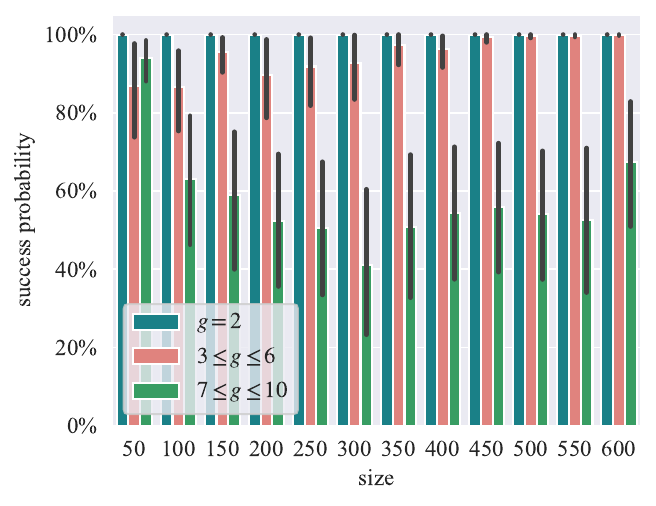}
    \caption{\HL{Success probabilities of the QTG-based search for instances of difficulties ranging from $g =2$ to $g =10$.}}
    \label{figure:QTGSuccessProbabilities}
\end{figure}

\HL{We carry out $100$ applications of the QTG-based \textbf{QMaxSearch} with a bias $b = n / 4$ and $M = 700 + n^{2} / 16$ Grover iterations.
As shown in \autoref{figure:QTGSuccessProbabilities}, the success probabilities of finding the optimal solution for instances with difficulty $2 \leq g \leq 6$ are above $80 \%$, while for hard instances with $7 \leq g \leq 10$ are above $40 \%$.
The quantum cycles and number of qubits required to run QTG were calculated as described in \autoref{section:Method} and averaged over all runs, see~\autoref{figure:QTGMemoryBenchmark} and \autoref{figure:QTGRuntimeBenchmark}.
We numerically observe that, for hard instances ($g \geq 7$), COMBO's cycle demand overtakes the quantum method's elapsed cycles already at instances with $100$ variables and indicates a steeper slope overall.}

\HL{Additionally, we observe that the QTG-based method requires far less memory than COMBO (assuming the comparability of bits and qubits) and any other tested method.
While the number of involved qubits stays constant for instances of the same size, COMBO's memory requirements increase exponentially for harder instances.}

\section{\label{section:DiscussionAndConclusion}Discussion and Conclusions}
\HL{In this paper, we have proposed and investigated the Quantum Tree Generator (QTG), a quantum method for solving instances of the 0-1 knapsack problem.
The QTG-based search emerges as a highly problem-specific version of quantum amplitude amplification with improved incorporation of the underlying classical structure at hand.
Due to our novel runtime calculator, we were able to collect numerical data, rather than mere extrapolation, for benchmark instances of various difficulty with up to 600 variables.}

\HL{On the evaluated benchmarks, the quantum method admitted favourable runtimes for hard instances with at least 100 variables while requiring a significantly smaller amount of memory.
Both results together -- runtime and (enormous) memory savings -- hint at a possible \emph{practical} quantum advantage for solving 0-1-$\KP$ instances.}

It is essential to acknowledge that there exist several other potential ways for improvement.
These areas of potential enhancement encompass the utilisation of upper bound heuristics within the QTG to reduce branching, or the utilisation of a preprocessing branch-and-bound procedure, such that the QTG-based search is executed only on the remaining subsequent problem, or to use a different bias function based on the weight of the items.

\section*{\label{section:Acknowledgments}Acknowledgments}
This work was supported by the DFG through SFB 1227 (DQ-mat), Quantum Frontiers, the Quantum Valley Lower Saxony, the BMBF projects ATIQ and QuBRA, and the BMWK project ProvideQ.
Helpful correspondence and discussions with Arne-Christian Voigt, Mark Bennemann, and Timo Ziegler are gratefully acknowledged.

\section*{\label{section:Authors}Author Contributions}
This project was conceived of, and initiated in, discussions of S.W. and T.J.O. The quantum implementation along with the resource estimation procedures were developed by S.W, L.B and A.I.L. The classical expertise was provided by S.F. and M.P. All authors contributed to writing the paper.

\twocolumngrid

\bibliographystyle{apsrev4-2}
\bibliography{main.bib}

\onecolumngrid
\clearpage

\twocolumngrid
\appendix

\section{\label{section:SupplementaryInformation}Supplementary Information}
\subsection{\label{subsubsection:ClassicalMethods}\HL{Classical Methods}}

\begin{figure}[h!]
    \centering
    \includegraphics[width=\linewidth]{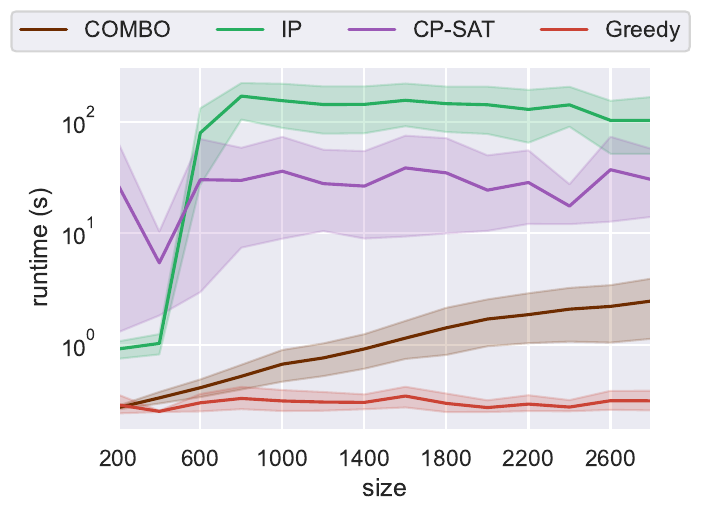}
    \caption{\HL{Comparison of several classical solvers on the \emph{jooken-large} instance set.
    We observe that despite COMBO being a single-threaded algorithm, it still outperforms the exact solvers Gurobi and CP-SAT in terms of runtime. Note that GUROBI and CP-SAT were not able to prove optimality for all instances within $300 \second$ despite finding an optimal solution for all of them.}}
    \label{figure:ComboVSGurobi}
\end{figure}

\HL{To compare the QTG-based search with classical methods, we evaluated the capabilities of classical knapsack solvers on a wide range of benchmark instances.
We use three exact solvers: GUROBI (based on IP), CP-SAT, and COMBO. 
EXPKNAP was also tested, but did not produce results within the given time frame and was therefore excluded from this evaluation.
Moreover, we also evaluated the performance of Greedy.
The instance set consisted of $336$ instances with $200-2800$ items that were generated with the algorithm proposed by \citet{Jooken2022ANewClassOfHardProblemInstancesForThe0_1KnapsackProblem}.
The experiment was carried out with a time limit of $300 \second$.}

\HL{\autoref{figure:ComboVSGurobi} presents a comparison between all classical algorithms. 
Note that although the simplicity of the Greedy algorithm allows for a short runtime, it produces suboptimal solutions in most cases.
On the other hand, IP and CP-SAT were able to find optimal solutions (within a gap of $10^{-5}$) for all instances. 
In some cases, however, both solvers were unable to prove optimality for found solutions resulting in exceeding the given time limit.
It is clear that COMBO being provably optimal and several orders faster than its exact counterparts, while only utilising a single core instead of the available $8$ cores serves as the basis for our comparison to QTG.}

\subsection{\label{subsection:BranchingStrategy}Branching strategy}

\subsubsection{\label{subsubsection:BiasedHadamardGates}Biased Hadamard Gates}

First we consider the biased Hadamard gates \eqref{equation:biasedHadamard} and note that, for $b = 0$, both variants yield the usual (unbiased) Hadamard gate H.
This would correspond to including or excluding the items with equal probability.
Accordingly, we numerically observed that the number of elapsed quantum cycles for obtaining a reasonable success probability would generally not be competitive when using the usual Hadamard gates.

In comparison, allowing for biased Hadamard gates and fine-tuning the bias $b$ can lead to an enormous speed-up during the phase of amplitude amplification:
Given an intermediate solution $\bm{x}'$, whose profit also constitutes the current threshold for the \textbf{QMaxSearch}, we can actively favour feasible paths $\bm{x}$ that have a certain Hamming distance $\hamming \coloneqq \hamming(\bm{x}', \bm{x})$ to $\bm{x}$.
Namely, applying the QTG and its biased Hadamard gates with arbitrary but fixed bias $b$, we can give the following lower bound on the sampling probability of $\bm{x}$:
\begin{linenomath}
\begin{align}\label{equation:biasedSamplingProbability}
    \lvert a_{\bm{x}}\rvert^{2} = \left(\frac{b + 1}{b + 2}\right)^{n - \hamming}\left(\frac{1}{b + 2}\right)^{\hamming}.
\end{align}
\end{linenomath}
This bound is tight if all possible branchings occur when constructing the path $\bm{x}$.
For each bit in which $\bm{x}$ and $\bm{x}'$ differ, the application of a biased Hadamard gate introduces the prefactor $1 / (b + 2)$. For each of the remaining bits, it establishes the prefactor $(b + 1) / (b + 2)$.

Now, we can determine $b$ in order to maximise the sampling probability of $\bm{x}$ within the superposition created by the QTG.
Differentiating \eqref{equation:biasedSamplingProbability} with respect to $b$ and checking for zeros yield the optimal bias
\begin{linenomath}
\begin{align*}
    b_{\opt} = \frac{n}{\hamming} - 2 \approx \frac{n}{\hamming}.
\end{align*}
\end{linenomath}

Fine-tuning the bias in order to sample paths with a certain hamming distance more frequently helps searching the neighbourhood of good intermediate solutions for even better solutions.
We observe that for many Knapsack instances the optimal solution and the Greedy result do have a lot of common bit values, corresponding to items whose efficiency is just too good to not be included in any optimal assignment.
Because we employ Greedy for obtaining an initial solution in our \textbf{QMaxSearch} routine, the biased Hadamard approach typically results in highly improved performance.

\HL{The time required by $\textbf{QSearch}$ algorithm to find a state with hamming distance $\hamming$ to a known solution is $\mathcal{O}(1 / a_x)$, where}
\begin{linenomath}
\HL{\begin{align}\label{eq:runtime}
\begin{aligned}
1 / a_x &=
    \left(\frac{b_{\opt} + 2}{b_{\opt} + 1}\right)^{(n - \hamming) / 2}\left(b_{\opt} + 2\right)^{\hamming / 2} \\
    &= \left(\frac{n}{n - \hamming}\right)^{(n - \hamming) / 2}\left(\frac{n}{\hamming}\right)^{\hamming / 2}    \approx \left(\frac{n}{\hamming}\right)^{\hamming / 2} .
\end{aligned}
\end{align}}
\end{linenomath}
\HL{More explicitly, we get a runtime $\mathcal{O}\left(\left(\frac{n}{\hamming}\right)^{\hamming / 2}\right)$.
As the neighbourhood of distance $\hamming$ to a known solution contains $\binom{n}{\hamming}\approx n^{\hamming}$ assignments, this runtime can be seen as a quadratic improvement over local search algorithms.}

\subsubsection{\label{subsubsection:UConstructionOfTheLayerUnitaries}Construction of the layer unitaries}

As described in \autoref{section:Method}, the QTG is composed of the layer unitaries $U_{m}$: $\QTG = \prod_{m = 1}^{n} U_{m}$.
That is, each $U_{m}$ manages the inclusion/exclusion of an item $m$.
The unitaries $U_{m}$ are themselves comprised of three components:
\begin{itemize}
    \item[1.] Create a biased superposition of including or excluding item $m$ if and only if including the item does not exceeds the capacity $c$, otherwise do nothing.
    We achieve this step by applying a multi-controlled biased Hadamard gate to the $m$-th qubit.
    The Hadamard gate itself acts on the $m$-th qubit in the register $\hil_{1}$.
    The control runs over the qubits of the second register $\hil_{2}$ and checks whether the remaining capacities cover the weight $w_{m}$ of including item $m$.
    Accordingly, we denote the control with $C_{\geq w_{m}}^{2}$, because it asks whether the second register's state, interpreted as the binary representation of an integer, is greater or equal to the integer $w_{m}$.
    It then highly depends on the actual value of $w_{m}$ which qubits in $\hil_{2}$ have to be involved into the control (see \autoref{subsubsection:Comparison}).
    In summary, this subgate is given by \HL{$U^{1}_{m} \coloneqq \controlled^{2}_{\geq w_{m}} (\text{H}^{x_{m}}_{b})$.}
    \item[2.] Update the remaining capacity.
    For this, we have to perform an integer subtraction by $w_{m}$ in the second register $\hil_{2}$ -- denoted by $\text{SUB}_{w_{m}}$ -- which is, however, only executed if the main register $\hil_{1}$ represents a path for which the $m$-th item is included.
    That is, the integer subtraction by $w_{m}$ is controlled by the $m$-th qubit in the main register.
    The entire update step therefore reads $U^{2}_{m} \coloneqq \controlled^{1}_{m} \text{SUB}_{w_{m}}$.
    \item[3.] Update the total profit.
    Here, we perform an integer addition by $p_{m}$ in the third register $\hil_{3}$ -- denoted by $\text{ADD}_{p_{m}}$ -- which is, analogously to step 2, controlled by the $m$-th qubit in the main register.
    This last update step thus reads $U^{3}_{m} \coloneqq \controlled^{1}_{m} \text{ADD}_{p_{m}}$.
\end{itemize}

To give an explicit example:
Let $\bm{x}$ be a feasible path whose $m$-th bit is set to zero and let $\tilde{\bm{x}}$ be the same bit string, but with its $m$-th bit set to one.
Further assume that the $m$-th bit of the intermediate solution $\bm{x}'$ has the value one.
The action of $U_{m}$ on a the feasible product state $\ket{\bm{x}}^{1} \ket{c_{\bm{x}}}^{2} \ket{P_{\bm{x}}}^{3}$ then reads

\begin{linenomath}
\HL{\begin{widetext}
\begin{equation}\label{equation:StatePreparationCircuitUnitary}
     U_{m} \ket{\bm{x}}^{1} \ket{c_{\bm{x}}}^{2} \ket{P_{\bm{x}}}^{3} = 
     \begin{cases}
        \ket{\bm{x}}^{1} \ket{c_{\bm{x}}}^{2} \ket{P_{\bm{x}}}^{3}, &\text{if } c_{\bm{x}} < w_{m} \\ \\
        \sqrt{\frac{1}{b + 2}} 
        \ket{\bm{x}}^{1} \ket{c_{\bm{x}}}^{2} \ket{P_{\bm{x}}}^{3} + \sqrt{\frac{b + 1}{b + 2}}
        \ket{\tilde{\bm{x}}}^{1} \ket{c_{\bm{x}} - w_{m}}^{2} \ket{P_{\bm{x}} + p_{m}}^{3}
        , &\text{else.}
    \end{cases}
\end{equation}
\end{widetext}}
\end{linenomath}

\subsection{\label{subsection:QuantumMaximumFinding}Quantum maximum finding}

Here, we recap the known quantum algorithms Amplitude Amplification/\textbf{QSearch} and \textbf{QMaxSearch} and detail our adjustments for application to the 0-1-KP.

\textbf{QSearch} is a direct generalisation of Grover's famous algorithm \cite{Grover1996AFastQuantumMechanicalAlgorithmForDatabaseSearch}.
More precisely, Grover's algorithm was initially proposed for finding a unique marked state within a uniform superposition.
It was then generalised to instances with multiple marked states in the form of the \emph{quantum exponential searching algorithm} \cite{Boyer1998TightBoundsOnQuantumSearching}.
The two central building blocks of these algorithms are the \emph{phase oracle}
\begin{linenomath}
\begin{align}\label{equation:PhaseOracle}
    \signflip_{f} \ket{\bm{x}} \coloneqq (-1)^{f(\bm{x})} \ket{\bm{x}},
\end{align}
\end{linenomath}
which flips the phase of all states corresponding to bit strings $\bm{x}$ that fulfil the Boolean trait $f(\bm{x}) = 1$, and the \emph{diffusion operator}
\begin{linenomath}
\begin{align}\label{equation:GroverDiffusionOperator}
    D = \text{H}^{\otimes n} \signflip_{0} \text{H}^{\otimes n}.
\end{align}
\end{linenomath}

In the more general \textbf{QSearch} routine, $\text{H}^{\otimes n}$ -- which creates a uniform superposition of all computational basis states -- is replaced by any other superposition-creating unitary $\mathcal{A}$, in our case by the QTG $\QTG$.
Fixing the Boolean function $f$ and the state preparation unitary $\mathcal{A}$, the Amplitude Amplification operator is given by
\begin{linenomath}
\begin{align}\label{equation:AmplitudeAmplificationOperator}
    \QSop \coloneqq \mathcal{A} \signflip_{0} \mathcal{A}^{\dagger} \signflip_{f}.
\end{align}
\end{linenomath}

In the original proposal \cite{Brassard2002QuantumAmplitudeAmplificationAndEstimation},
the \textbf{QSearch} algorithm runs forever if there is no state $\bm{x}$ fulfilling $f(\bm{x})$.
In our case, however, this situation naturally occurs at the end of an entire \textbf{QMaxSearch} routine (see remainder of this section) which is why we additionally introduce a cutoff to prevent an endless loop.
With this extension, our version of \textbf{QSearch} reads

\begin{algorithm}[!ht]
\caption{\textbf{QSearch} ($\QTG$, $f$, $M$)}
Set $l = 0$, $m_{\total} = 0$, and let $d$ be any constant such that $1 < d < 2$\;
Increase $l$ by $1$ and set $m = \lceil d^{l}\rceil$\;
Apply $\QTG$ to the initial state $\ket{\bm{0}}^{1} \ket{c}^{2} \ket{0}^{3}$\;
Choose $j \in [1, m]$ uniformly at random and increase $m_{\total}$ by $2 j + 1$\;
Apply $\QSop^{j}$ to the superposition $\ket{s}^{1 2 3} = \QTG \ket{\bm{0}}^{1} \ket{c}^{2} \ket{0}^{3}$\;
Measure the register A and C, and get the outcome $\ket{\bm{x}}^{1}
\ket{P_{\bm{x}}}^{3}$\;
\eIf{$f(P_{\bm{x}}) = 1$ or $m_{\total} \geq M$}{
    return $\bm{x}$, $P_{\bm{x}}$\;
}{
    go to step 2\;
}
\end{algorithm}

Because we are interested in maximising the total profit among all feasible paths, we have to choose the Boolean function $f$ accordingly.
The apt method in question is provided by Quantum Maximum Finding.
While initially proposed as an extension to Grover's algorithm, its principle is directly transferable to \textbf{QSearch}.

The key idea is to introduce the comparison ``$P_{\bm{x}} > T$'' as Boolean function which marks all paths $\bm{x}$ whose total profit $P_{\bm{x}}$ exceeds a given threshold $T$.
Starting with an initial threshold $T_{0}$, \textbf{QSearch}$(\QTG, P_{\bm{x}} > T_{0}, M)$ yields an intermediate solution $(\bm{x}_{1}, P_{\bm{x}_{1}})$.
If $P_{\bm{x}_{1}} > T_{0}$, we can call \textbf{QSearch} again with the updated threshold $T_{1} \coloneqq P_{\bm{x}_{1}}$ and so on.
However, if we obtain an output $(\bm{x}_{i}, P_{\bm{x}_{i}})$ with $P_{\bm{x}_{i}} \leq T_{i - 1}$ throughout this routine, we can conclude -- within a certain probability -- that no valid path exists with a higher total profit than $T_{i - 1}$;
hence we declare the previous intermediate solution $(\bm{x}_{i - 1}, T_{i - 1})$ the optimal solution.
Because this case eventually occurs for every instance, we have introduced the additional termination condition in \textbf{QSearch}.
Finally, for the initial threshold we employ the very efficient Integer Greedy method.

\begin{algorithm}[!ht]
\caption{\textbf{QMaxSearch}($\QTG$, $M$, $\bm{p}$, $\bm{w}$, $c$)}
Calculate initial threshold $T$ using Greedy:  
$\bm{x}_{\out}$, $T = \textbf{intGr}(\bm{p}, \bm{w}, c)$ \;
\While{True}{
$\bm{x}$, $P_{\bm{x}}$ = \textbf{QSearch}($\QTG$, $P_{\bm{x}} > T$, $M$)\;
\eIf{$P_{\bm{x}} > T$}{
    set $T = P_{\bm{x}}$ and $\bm{x}_{\out} = \bm{x}$\;
}{
    return $\bm{x}_{\out}$, $T$\;
}
}
\end{algorithm}

\subsection{\label{subsection:RuntimeCalculator}\HL{Runtime calculator}}

\HL{The instance sizes considered in \autoref{figure:QTGMemoryBenchmark} and \autoref{figure:QTGRuntimeBenchmark} are way too large for any conventional simulator which is based on evaluating the action of general quantum gates on general quantum states.
Even with the most sophisticated simulation technique and with the aid of the largest available super computer, these approaches could only tackle up to 49 qubits\cite{Smelyanskiy2016qHiPSTERTheQuantumHighPerformanceSoftwareTestingEnvironment}.
Instead, we developed a completely different runtime calculator tailored explicitly to the application of the QTG and \textbf{QMaxSearch}.
The calculator consists of two main parts:
\begin{enumerate}
    \item calculate the QTG's action only on computational basis states whose objective value surpasses the current threshold imposed by \textbf{QMaxSearch}.
    \item calculate the effect of Amplitude Amplification operator via analytical formulas.
\end{enumerate}}

\subsubsection{\label{subsubsection:QTGAsAStateSieve}\HL{QTG as a state sieve}}

We classically mimic the QTG by exploring the classical binary decision tree for a given knapsack instance using \textit{breadth-first search} (BFS) and, in doing so, keep track of the remaining capacities, the total profit, and the sampling probabilities of the computational basis states after each application of one layer operator $U_{m}$.
The sampling probabilities are propagated as follows:
Assume a node in the $m$-th layer carries a remaining capacity of $\tilde{c}$, a total profit of $p$, and a sampling probability of $q$.
If $w_{m + 1} > \tilde{c}$, no branching occurs and the only child node inherits its data from its parent node.
If instead $w_{m + 1} \leq \tilde{c}$, a branching produces two child nodes corresponding to either excluding (left) or including (right) the $(m + 1)$-th item.
The child nodes' metadata are:
\begin{linenomath}
\begin{align*}
    (\tilde{c}_{\text{left}}, \tilde{c}_{\text{right}}) &= (\tilde{c}, \tilde{c} - w_{m + 1}) \\
    (p_{\text{left}}, p_{\text{right}}) &= (p, p + p_{m + 1}) \\
    (q_{\text{left}}, q_{\text{right}}) &= \begin{cases}
        \big(\frac{q (b + 1)}{b + 2}, \frac{q}{b + 2}\big), & \text{if } x_{m + 1} = 0 \\
        \big(\frac{q}{b + 2}, \frac{q (b + 1)}{b + 2}\big), & \text{else.}
    \end{cases}
\end{align*}
\end{linenomath}
The update rule for the child nodes' sampling probability directly results from the definition of the biased Hadamard gate \eqref{equation:biasedHadamard}.

During the BFS, we prune infeasible subtrees as the QTG would do anyway on a real quantum device, thereby massively reducing the final number of states.
Furthermore, we also discard subtrees which do not admit any path with an objective value surpassing the current threshold.
For this, we execute COMBO on the Knapsack subinstance that remains when fixing the first $m$ decision variables \HL{$x_{1}, \ldots, x_{m}$} with the values corresponding to the current node in the decision tree.
This yields a tight upper bound for the optimal profit achievable within the respective subtree.
This second pruning step, although computationally sumptuous, reduces the final number of states in such a drastic way that even for \HL{large instances up to 600 variables, only a few million -- instead of $2^{600}$ --  states remain.}

The result of this runtime calculator, given a threshold $T$, is an array of paths (bit strings) together with their remaining capacity, total profit, and the theoretical sampling probability of the associated computational basis state.
All these paths are feasible and admit, by construction, a respective total profit above the threshold $T$.
This final array of states now serves as the input for Amplitude Amplification.

We denote this set of paths by $\mathcal{F}^{T}$ and have therefore gained a representation of the superposition
\begin{linenomath}
\HL{\begin{equation}
\begin{aligned}
    \ket{s}^{1 2 3} &= \QTG \ket{\bm{0}}^{1} \ket{c}^{2} \ket{0}^{3} \\
    &= \sum_{\bm{x} \scriptin \mathcal{F}^{T}} a_{\bm{x}} \ket{\bm{x}}^{1} \ket{c_{\bm{x}}}^{2} \ket{P_{\bm{x}}}^{3} + a_{R} \ket{R} \\
    &= a_{T} \ket{\mathcal{F}^{T}} + a_{R} \ket{R},
\end{aligned}
\end{equation}}
\end{linenomath}
where
\begin{linenomath}
\begin{align*}
    \abs{a_{T}}^{2} \coloneqq \sum_{\bm{x} \scriptin \mathcal{F}^{T}} \abs{a_{\bm{x}}}^{2}\text{ and } \abs{a_{R}}^{2} = 1 - \abs{a_{T}}^{2}.
\end{align*}
\end{linenomath}
All other feasible paths, but with a total profit below or equal to $T$, are thus collectively represented by the remainder state $\ket{R}$.
In most cases, the remainder state's amplitude is nonzero, such that the cumulative sampling probability $q_{T} \coloneqq \abs{a_{T}}^{2}$ of all the paths in $\mathcal{F}^{T}$ is not given by one.

\begin{figure*}[!ht]
    \centering
    \begin{minipage}{0.51\textwidth}
        \centering
        \includegraphics[height=0.125\textheight]{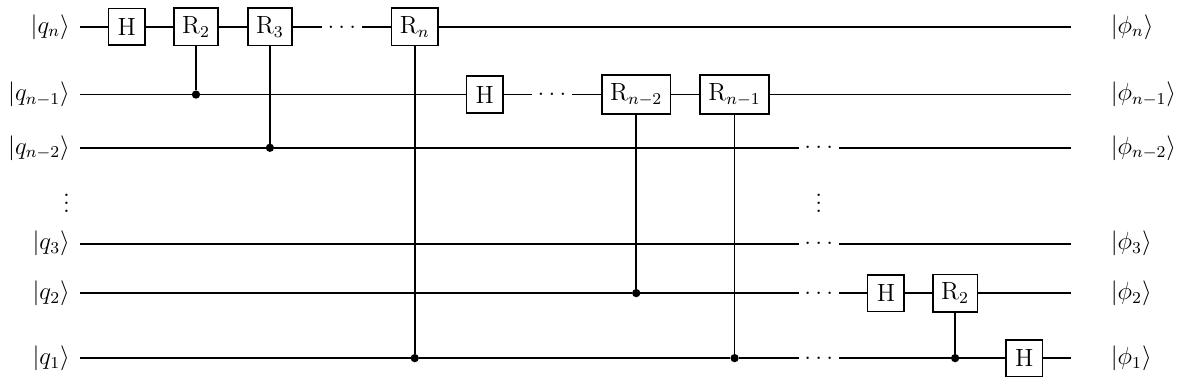}
    \end{minipage}\hfill
    \begin{minipage}{0.49\textwidth}
        \includegraphics[height=0.135\textheight]{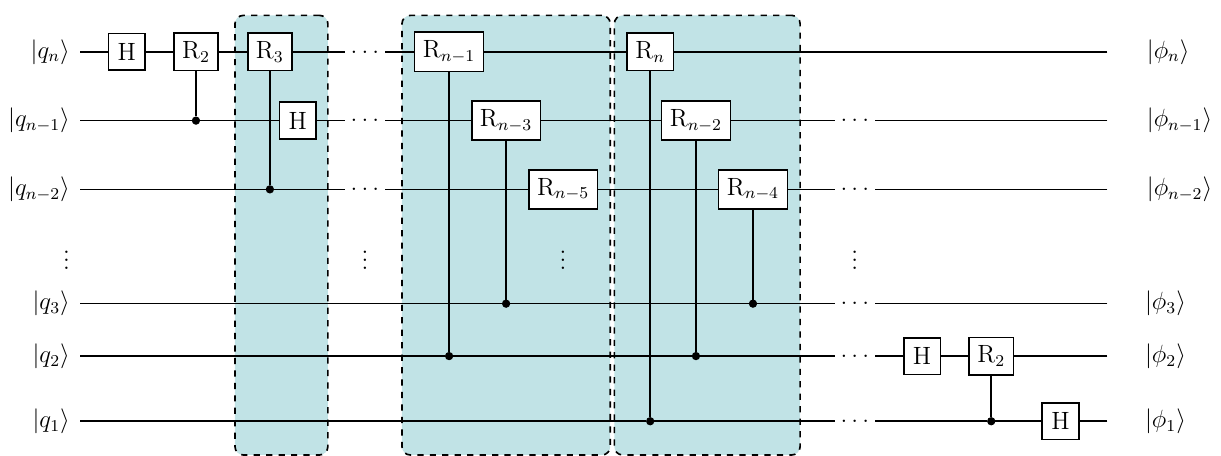}
    \end{minipage}
        \caption{\label{figure:QFTCircuit}
            \textbf{a)} The standard, non-parallelised QFT circuit.
            \textbf{b)} The optimally parallelised QFT circuit.
            For all $i = 3, \ldots, n$, the gates in the $\ket{q_{i - 1}}$-level are shifted to the left until the leftmost gate, the Hadamard gate, is executed in parallel with the R$_{3}$ gate in the $\ket{q_{i}}$-level.
            As the $\ket{q_{2}}$-level does not admit an R$_{3}$ gate, the Hadamard gate in the $\ket{q_{1}}$-level is executed alone.
            Therefore, for all $i = 2, \ldots, n$, the $\ket{q_{i - 1}}$-level gate sequence with its $i - 1$ gates starts two layers after the $\ket{q_{i}}$-level's sequence which is comprised of $i$ gates.
            Starting from the $\ket{q_{n}}$-level, each additional level thus introduces only one additional layer to the circuit.
            }
\end{figure*}

\subsubsection{\label{subsubsection:AmplitudeAmplification}Amplitude Amplification}

Beside the QTG, we also have to account for the action of the Amplitude Amplification operator $\QSop$ (step 5 in \textbf{QSearch}) and quantum measurements (step 6 in \textbf{QSearch}).
Choosing some power $j \in \N$, the action of $\QSop^{j}$ on $\ket{s}^{1 2 3}$ is explicitly given\cite{Brassard2002QuantumAmplitudeAmplificationAndEstimation} by
\begin{linenomath}
\begin{align}\label{equation:AmplitudeAmlplifcationAction}
\begin{split}
    \QSop^{j} \ket{s}^{1 2 3} =\ &\frac{\sin\left((2 j + 1)\arcsin\left(\sqrt{q_{T}}\right)\right)}{\sqrt{q_{T}}} \ket{\mathcal{F}^{T}} \\
    &+ \frac{\cos\left((2 j + 1) \arcsin\left(\sqrt{q_{T}}\right)\right)}{\sqrt{1 - q_{T}}} \ket{R}.
\end{split}
\end{align}
\end{linenomath}

The crucial aspect lies in the amplification of the amplitudes of the all the states contributing to the superposition $\ket{\mathcal{F}^{T}}$ -- although being nonuniformly distributed -- by the same factor.
Therefore, the entire the action of $\QSop^{j}$ consists of merely multiplying each probability handed over by the QTG sieve with the common amplification factor
\begin{linenomath}
\begin{align}\label{equation:AmplitudeAmplificationFactor}
    \frac{\sin^{2}\left((2 j + 1) \arcsin\left(\sqrt{q_{T}}\right)\right)}{q_{T}}.
\end{align}
\end{linenomath}

Finally, simulating a quantum measurement in this framework is straightforward:
Uniformly draw a random number from the interval $[0, 1]$ (we use GNU Scientific Library's implementation of the MT19937 algorithm \cite{Matsumoto1998MersenneTwisterA623DimensionallyEquidistributedUniformPseudoRandomNumberGenerator}).
Then iteratively sum up the sampling probabilities of all the paths in $\mathcal{F}^{T}$ until the cumulative probability surpasses the random number.
If this occurs before the loop ends, the last added path is considered to have been measured and is returned (lines 7 and 8 in \textbf{QSearch}).
Otherwise, that is if the random number exceeds $q_{T}$, we consider the remainder state being sampled (lines 9 and 10 in \textbf{QSearch}).
Because in the latter case, we simply omit the result and restart the entire procedure with a new value for $j$, no further information is necessary.
This completely justifies our approach of neglecting all the information about feasible states that admit a total profit below the given threshold.

\subsection{\label{subsection:ResourceEstimation}Resource estimation}

This section contains detailed resource estimations for the subcircuits required for implementing the QTG-based search, specifically, the costs associated with implementing the exploration unitaries $U_{m} = U^{3}_{m} U^{2}_{m} U^{1}_{m}$, $m = 1, \ldots, n$.
A crucial aspect of evaluating the efficiency of our quantum algorithm is to determine the total number of qubits $\qubitc$, gates $\gatec$, and cycles $\cyclec$ required for its implementation.
Here, a cycle refers to a collection of gates that can be executed in parallel.
In the context of resource estimation, we make the following assumptions:
\begin{enumerate}
    \item[\textbf{(i)}] Any $n$-controlled single-qubit gate can be decomposed into $2 (n - 1)$ Toffoli gates and a single-controlled single-qubit gate.
    This decomposition utilises $n - 1$ ancilla qubits.
    \item[\textbf{(ii)}] All multi-controlled single-qubit gates have access to the same ancilla qubits arising from \textbf{(i)}.
    This corresponds to taking the maximum instead of the sum of all additionally introduced ancilla qubits.
    \item[\textbf{(iii)}] It is equally expensive to have a controlled gate on the logical state 0 or 1.
\end{enumerate}

\begin{figure*}[!ht]
    \centering
    \begin{minipage}{0.5\textwidth}
        \centering
        \includegraphics[height=0.17\textheight]{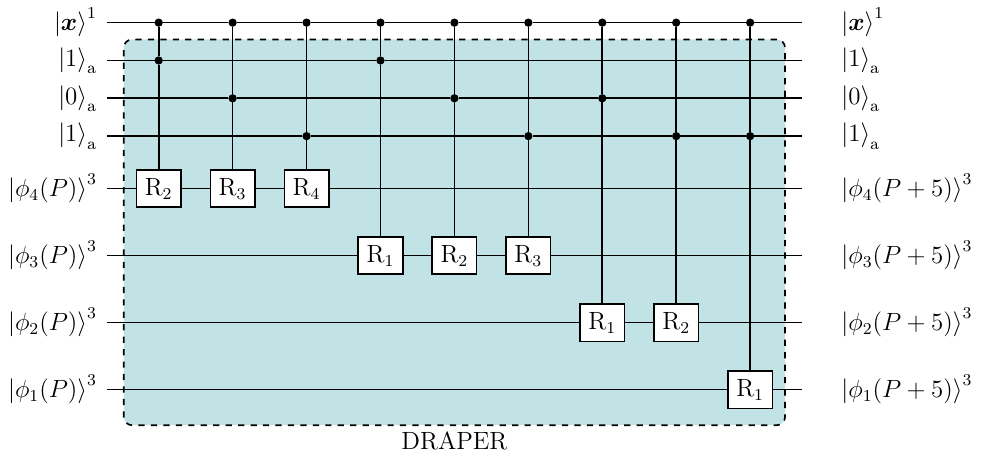}
    \end{minipage}\hfill%
    \begin{minipage}{0.5\textwidth}
        \centering
        \includegraphics[height=0.17\textheight]{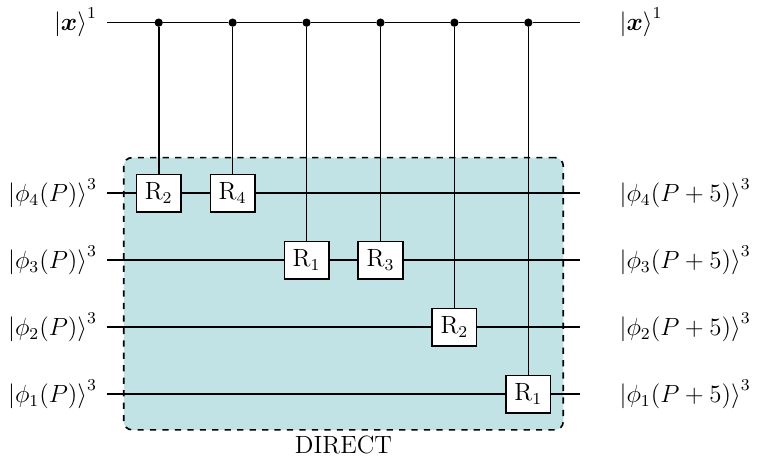}
    \end{minipage}
    \caption{\label{figure:AdditionMethods}
        Addition of the item profit $p_{m} = 5$ to the QFT-ed profit register, controlled on the $m$-th qubit in the path register.
        \textbf{a)} Draper method.
        The binary representation of the integer $5$ is encoded into the three-qubit ancilla state $\ket{101}_{\ancilla}$ whose qubits control the application of the rotations R$_{k}$ on the profit register.
        \textbf{b)} Direct method.
        The rotations for which the controlling ancilla qubit in Draper's method would be in the $\ket{1}_{\ancilla}$ state are directly implemented, the remaining ones simply omitted.
        In comparison to Draper's method, this saves several gates and the overall ancilla state preparation.
        }
\end{figure*}
\subsubsection{\label{subsubsection:QubitCount}Qubit count}

The total number of qubits is given by the sum of the main registers' sizes \eqref{equation:TotalNumberOfQubits} and the number of ancilla qubits for decomposing multi-controlled gates according to \textbf{(i)}.
Specifically, the size of the path register corresponds to the number of items involved in the Knapsack problem
\begin{linenomath}
\begin{equation}\label{equation:PathRegisterQubits}
    \qubitc_\text{PATH} = n.
\end{equation}
\end{linenomath}
The size of the capacity register is determined by the length of the capacity's binary representation, i.e.,
\begin{linenomath}
\HL{\begin{equation}\label{equation:CapacityRegisterQubits}
    \qubitc_{\text{CAPACITY}} = \numbits{c}.
\end{equation}}
\end{linenomath}
Analogously, given an upper bound $P$ for the optimal profit, the size of the profit register is given by the length of $P$'s binary representation, that is
\begin{linenomath}
\HL{\begin{equation}\label{equation:ProfitRegisterQubits}
    \qubitc_{\text{PROFIT}} =  \numbits{P}.
\end{equation}}
\end{linenomath}
Additionally, the QTG contains the comparisons ``$\tilde{c} \geq w_{m}$'' for all the item weights $w_{m}$, $m = 1, \ldots, n$, which constitute multi-controlled single-qubit gates (see \autoref{subsubsection:Comparison}).
Applying \textbf{(i)}, the decomposition of those gates introduces up to $\numbits{c} - 1$ ancilla qubits.
For simplicity, we assume that this upper bound is actually achieved by at least one of the comparisons.
Due to assumption \textbf{(ii)}, this number of ancilla qubits then suffices to implement all the comparisons.
Analogously, the Amplitude Amplification operator $\QSop$ contains the phase oracles $\signflip_{0}$ and $\signflip_{P_{\bm{x} > T}}$ which we realise using $n$-controlled and $\numbits{P}$-controlled single-qubit gates, respectively (see \autoref{subsubsection:Comparison}).
Hence, following \textbf{(i)} and \textbf{(ii)}, \HL{$\max(n, \numbits{c}, \numbits{P}) - 1$ additional ancilla qubits have to be introduced}.
Utilising \textbf{(ii)} again, these ancilla qubits then even suffice to execute $\QSop$ several times with different thresholds $T$.
\HL{Lastly, we use one additional qubit to store the potential phase flip applied by $\QSop$.
In total, we obtain an amount of}
\begin{linenomath}
\HL{\begin{align}\label{equation:AncillaQubits}
  \qubitc_\text{ANC} &= \max(n, \numbits{c}, \numbits{P})
\end{align}}
\end{linenomath}
\HL{ancilla qubits.
In summary, our methods requires the following number of qubits:}
\begin{linenomath}
\HL{\begin{align}\label{equation:TotalQubits}
    \qubitc_{\text{TOTAL}} &= \qubitc_{\text{PATH}} + \qubitc_{\text{CAPACITY}} + \qubitc_{\text{PROFIT}} + \qubitc_{\text{ANC}} \nonumber \\
    &= n + \numbits{c} + \numbits{P} \nonumber \\
    &\quad + \max(n, \numbits{c}, \numbits{P}).
\end{align}}
\end{linenomath}

\begin{figure}
    \centering
    \includegraphics[scale=0.6]{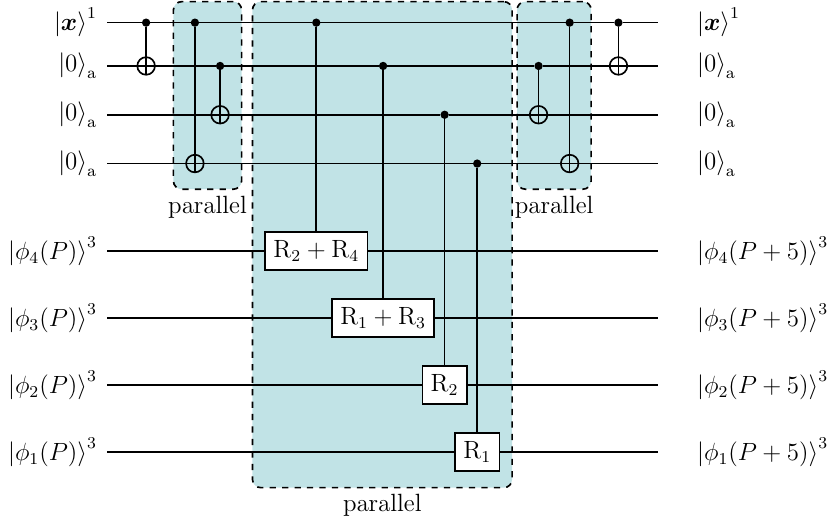}
    \caption{\HL{Optimised parallelisation of addition.
        The consecutive rotations gates, controlled by the same qubits, can be further merged into a single rotation gate. By introducing 3 ancilla qubits, we can further execute the addition gates within a single cycle. This comes at the expense of requiring four additional cycles that are given by duplicating the path register.}}
    \label{figure:OptimisedAddition}
\end{figure}

\subsubsection{\label{subsubsection:QFT}QFT}

\begin{figure*}[!ht]
    \centering
    \begin{minipage}{.47\textwidth}
        \centering
        \includegraphics[height=0.15\textheight]{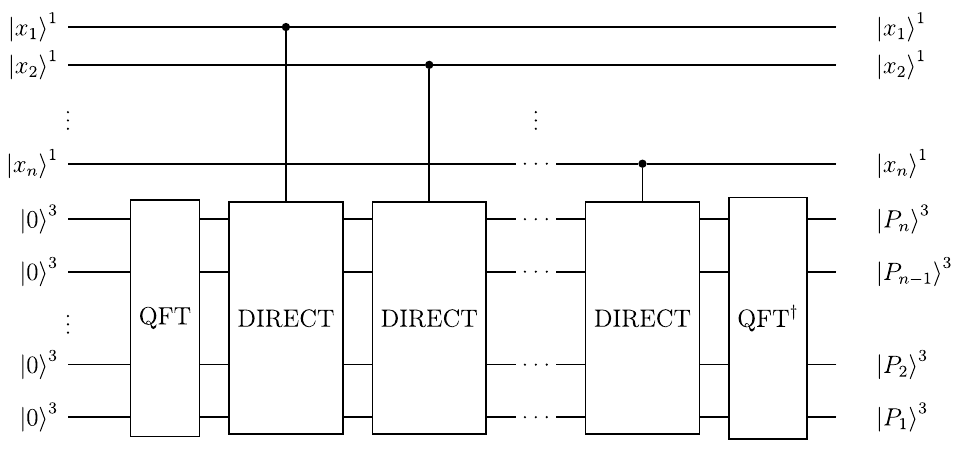}
    \end{minipage}\hfill%
    \begin{minipage}{0.52\textwidth}
        \centering
        \includegraphics[height=0.15\textheight]{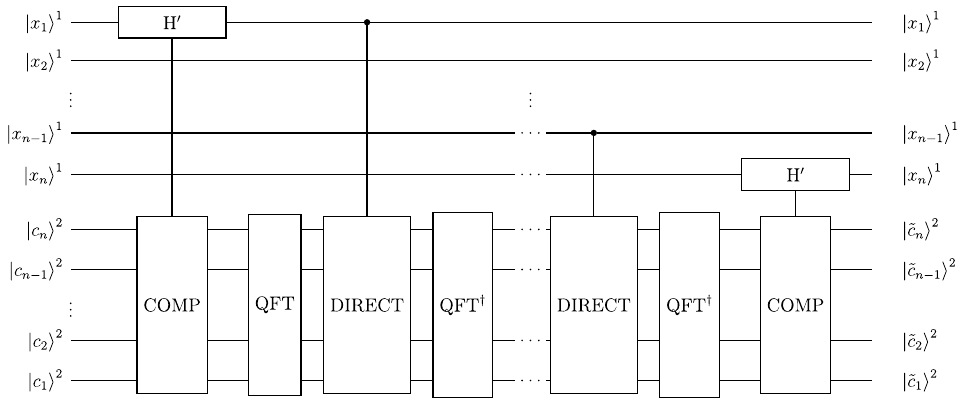}
    \end{minipage}
    \caption{\label{figure:AdditionSubtractionCircuits}
        \textbf{a)} The subcircuit performing all the profit additions.
        The profit register is QFT-ed in the beginning.
        Subsequently, the respective item profits are added via the direct method.
        The controlled Hadamard gates and the weight subtraction between the single additions are not depicted.
        After all item profits are added, the register is transformed back.
        \textbf{b)} The subtraction circuit including the comparison part.
        The biased Hadamard gates on the path register are controlled on the capacity register by integer comparisons.
        After the comparison the item weights are subtracted from the capacity register, controlled on the qubit just affected by the biased Hadamard gate.
        Because of this alternating involvement of the capacity register, the intermediate QFTs and their inverses do not cancel each other.
    }
\end{figure*}

Crucial for the efficient implementation of quantum adders (see \autoref{subsubsection:QFTAdder} and \autoref{subsubsection:QFTSubtractor}) is the well-known Quantum Fourier Transform \cite{Coppersmith2002AnApproximateFourierTransformUsefulInQuantumFactoring} (QFT).
\autoref{figure:QFTCircuit} shows the general circuit representation of the QFT -- unparallelised and parallelised -- with shorthand notation
\begin{linenomath}
\begin{align*}
    \text{R}_{k} \coloneqq \begin{bmatrix}
        1 & 0 \\
        0 & e^{2 \pi i / 2^{k}} \\
    \end{bmatrix}.
\end{align*}
\end{linenomath}

\paragraph*{Gate count.}
The number of gates involved in the QFT is fully determined by the underlying register's size.
Namely, to perform the QFT on the capacity register, the total number of gates required is given by
\begin{linenomath}
\HL{\begin{equation}\label{equation:CapacityQFTGates}
\gatec_{\text{QFT-CAPACITY}} = \dfrac{\numbits{c} (\numbits{c} + 1)}{2}.
\end{equation}}
\end{linenomath}
Similarly, for the profit register, the total number of gates needed for the QFT is
\begin{linenomath}
\HL{\begin{equation}\label{equation:ProfitQFTGates}
\gatec_{\text{QFT-PROFIT}} = \dfrac{\numbits{P} (\numbits{P} + 1)}{2}.
\end{equation}}
\end{linenomath}

\paragraph*{Cycle count.}
Following the parallelisation strategy depicted in \autoref{figure:QFTCircuit}, the number of cycles can be significantly reduced:
The level-specific gate sequences can be shifted such that each level merely introduces one additional cycle to the nonparallelisable gate sequence in the highest level.
As the length of the latter is given by the underlying register's size, we derive the following formulas for the optimised QFT cycle count:
\begin{linenomath}
\HL{\begin{align}
    \cyclec_{\text{QFT-CAPACITY}} &= 2 \numbits{c} - 1, \label{equation:CapacityQFTCycles} \\
    \cyclec_{\text{QFT-PROFIT}} &= 2 \numbits{P} - 1 \label{equation:ProfitQFTCycles}.
\end{align}}
\end{linenomath}

\subsubsection{\label{subsubsection:QFTAdder}QFT Adder}

Our version of the QFT adder is derived from the original proposal by Draper \cite{Draper2000AdditionOnAQuantumComputer}.
The latter is designed to add together two integers $n_{1}$ and $n_{2}$ which are encoded into quantum states $\ket{n_{1}}$ and $\ket{n_{2}}$ on two different registers.
The result $\ket{n_{1} + n_{2}}$ is then stored in, say, the first register.
The entire QFT adder circuit consists of applying the QFT to the first register, then performing a cascade of rotations on the first register, controlled on the qubits of the second register, and finally applying the inverse QFT to the first register (see \autoref{figure:AdditionMethods} for an example on the profit register with additional control on some qubit in the path register).
Because the items' profits that we wish to add are known in advance, we can instead remove the controls and implement all the rotations, controlled on qubits which would have been initialised in the $\ket{1}$, directly.
In a second optimisation step, we merge all the rotations acting on the same qubits, thus reducing the overall gate count.
Furthermore, by copying the control qubit's state multiple times on ancilla qubits, we substitute the shared control with controls on distinct qubits.
This allows us to entirely parallelise the the actual addition while introducing only logarithmically many additional copying cycles.
\autoref{figure:OptimisedAddition} shows an example of the direct addition method and its parallelised version which, in comparison to Draper's method, requires fewer gates and cycles.

\paragraph*{Gate count.}
The number of actually implemented gates in the addition step depends on the main register's size and the number we wish to add, more precisely, on the position of the least significant one in its binary representation:
In the original proposal of Draper, the $i$-th qubit in the summand's register controls $S - i + 1$ rotations in the main register, where $S$ is the size of the main register.
In the direct method, those $S - i + 1$ gates are implemented if and only if the $i$-th qubit is in the $\ket{1}$ state, that is if and only if the classical $i$-th bit is one.
We can see that the least significant one in the bit string introduces the most rotation gates.
As all rotations arising from higher bit positions are merged with the rotations stemming from this least significant one, already giving the number of rotations in the pure addition step.
The number of states copied beforehand requires one less qubit than the number of qubits acted on in the addition step;
the same amount of gates have to be applied after the addition in order to uncompute the ancilla qubits.

\HL{For a given profit $p_{m}$, let LSO$(p_{m})$ denote the index/position of its least significant one;
the gate count for adding $p_{m}$ to the QFT-ed profit register is then given by}
\begin{linenomath}
\HL{\begin{equation}\label{equation:AdditionGates}
    \gatec_{\text{DIRECT}_{p_{m}}} = 3 (\numbits{P} - \text{LSO}(p_{m})) + 1.
\end{equation}}
\end{linenomath}
Including the QFT and its inverse in the overall gate count for the additions, we obtain the following gate count for the QFT adder which adds the number $p_{m}$ to the profit register:
\begin{linenomath}
\HL{\begin{align}\label{equation:QFTAdderGates}
    \gatec_{\text{ADD}_{p_{m}}} &= 2 \gatec_{\text{QFT-PROFIT}} + \gatec_{\text{DIRECT}_{p_{m}}} \nonumber \\ 
    &= \numbits{P} (\numbits{P} + 1) \nonumber \\
    &\quad + 3 (\numbits{P} - \text{LSO}(p_{m})) + 1.
\end{align}}
\end{linenomath}

In the QTG circuit, all the additions on the profit registers are consecutive (see \autoref{figure:AdditionSubtractionCircuits}) which allows for cancelling the inverse QFT within ADD$_{p_{m}}$ and the QFT in ADD$_{p_{m + 1}}$ for all $m = 1, \ldots, n - 1$.
As a result, only one QFT and one inverse QFT are applied to the profit register in total.
The overall gate count for the addition of all item profits therefore reads
\begin{linenomath}
\HL{\begin{align}\label{equation:AllQFTAdderGates}
    \gatec_{\text{ADD}_{\bm{p}}} &= 2 \gatec_{\text{QFT-PROFIT}} + \sum_{m = 1}^{n} \gatec_{\text{DIRECT}_{p_{m}}} \nonumber \\
    &= \numbits{P} (\numbits{P} + 1) \nonumber \\
    &\quad + \sum_{m = 1}^{n} (3 (\numbits{P} - \text{LSO}(p_{m})) + 1).
\end{align}}
\end{linenomath}

\paragraph*{Cycle count.}
\HL{Due to the distributed control onto several ancilla qubits all rotational gates can be executed in parallel.
The control on the QFT and inverse QFT can be dropped completely.
The update of the ancilla qubits as well as the reverse operation can be parallelised into logarithmically many cycles, respectively.
In summary, we obtain here}
\begin{linenomath}
\HL{\begin{align}\label{equation:AllQFTAdderCycles}
    \cyclec_{\text{ADD}_{\bm{p}}} &= 2 \cyclec_{\text{QFT-PROFIT}} \nonumber \\
    &\quad + \sum_{m = 1}^{n} \big(2 \lceil\log_{2}(\numbits{P} - \text{LSO}(p_{m}))\rceil + 1\big) \nonumber \\
    &= 4 \numbits{P} - 2 \nonumber \\
    &\quad + \sum_{m = 1}^{n} \big(2 \lceil\log_{2}(\numbits{P} - \text{LSO}(p_{m}))\rceil + 1\big).
\end{align}}
\end{linenomath}

\subsubsection{\label{subsubsection:QFTSubtractor}QFT Subtractor}

By flipping all the rotations' signs, the QFT adder circuit can be used to subtract a given integer.
Therefore, adding and subtracting both yield the same gate count.

\paragraph*{Gate count.}
For a given profit $w_{m}$, let LSO$(w_{m})$ denote its least significant one's position;
the gate count for subtracting $w_{m}$ from the QFT-ed capacity register is then given by
\begin{linenomath}
\HL{\begin{equation}\label{equation:SubtractionGates}
    \gatec_{\text{DIRECT}_{w_{m}}} = 3 (\numbits{c} - \text{LSO}(w_{m})) + 1.
\end{equation}}
\end{linenomath}

Analogous to the QFT adder, including the QFT and its inverse in the overall gate count for the subtractions, we obtain the following gate count for the QFT subtractor which subtracts the number $w_{m}$ from the capacity register
\begin{linenomath}
\HL{\begin{align}\label{equation:QFTSubtractorGates}
    \gatec_{\text{SUB}_{w_{m}}} &= 2 \gatec_{\text{QFT-CAPACITY}} + \gatec_{\text{DIRECT}_{w_{m}}} \nonumber \\ 
    &= \numbits{c} (\numbits{c} + 1) \nonumber \\
    &\quad + 3 (\numbits{c} - \text{LSO}(w_{m})) + 1.
\end{align}}
\end{linenomath}

Unlike the profit additions, the weight subtractions in the QTG circuit are followed by a comparison procedure, respectively, hence the intermediate QFT and inverse QFT do not cancel each other (see \autoref{figure:AdditionSubtractionCircuits}).
To reduce the number of gates, however, we omit the last subtraction SUB$_{w_{n}}$, because the capacity register is not checked again after the last item inclusion.
All together, this leads to a total gate count of
\begin{linenomath}
\HL{\begin{align}\label{equation:AllQFTSubtractorGates}
    \gatec_{\text{SUB}_{\bm{w}}} &= 2 (n - 1) \gatec_{\text{QFT-CAPACITY}} + \sum_{m = 1}^{n - 1} \gatec_{\text{DIRECT}_{w_{m}}} \nonumber \\
    &= (n - 1) \numbits{c} (\numbits{c} + 1) \nonumber \\
    &\quad + \sum_{m = 1}^{n - 1} (3 (\numbits{c} - \text{LSO}(w_{m})) + 1).
\end{align}}
\end{linenomath}

\paragraph*{Cycle count.}
\HL{In analogy to the QFT adder, we obtain a total cycle count of}
\begin{linenomath}
\HL{\begin{align}\label{equation:AllQFTSubtractorCycles}
    \cyclec_{\text{SUB}_{\bm{w}}} &= 2 (n - 1) \cyclec_{\text{QFT-CAPACITY}} \nonumber \\
    &\quad + \sum_{m = 1}^{n - 1} \big(2 \lceil\log_{2}(\numbits{c} - \text{LSO}(w_{m}))\rceil + 1\big) \nonumber \\
    &= 4 (n - 1) \numbits{c} - 2 (n - 1) \nonumber \\
    &\quad + \sum_{m = 1}^{n - 1} \big(2 \lceil\log_{2}(\numbits{c} - \text{LSO}(w_{m}))\rceil + 1\big).
\end{align}}
\end{linenomath}

\subsubsection{\label{subsubsection:Comparison}Comparison}
\begin{figure}
    \centering
    \includegraphics[scale=0.64]{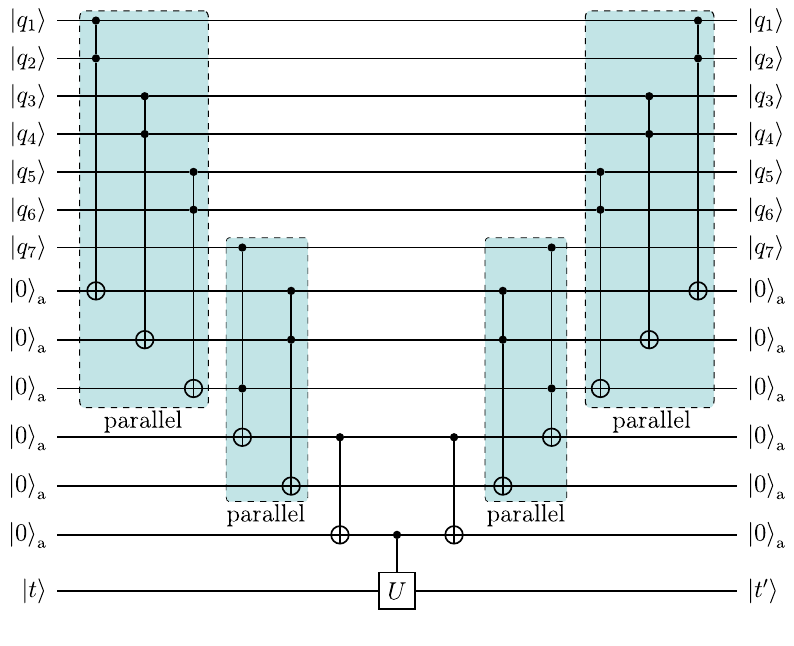}
    \caption{The optimal decomposition of a 7-qubit-controlled unitary $U$. In contrast to the standard decomposition method (see \cite{NielsenChuang2011QuantumComputationAndQuantumInformation})which lacks gate parallelisation, our approach reduces the cycle count from 13 to 7 .\label{figure:ParallelisedMultiControlDecomposition}}
\end{figure}

Within the QTG-based search, there are three subcircuits requiring the comparison of two integers:
First, the biased Hadamard gates within the QTG should only be executed if the remaining capacity in the second register is at least as large as the respective item weight.
Therefore, we have to implement the integer comparisons ``$\tilde{c} \geq w_{m}$'' with unknown $\tilde{c}$, but known $w_{m}$, $m = 1, \ldots, n - 1$.
Second, the Amplitude Amplification operator $\QSop$ contains the reflection $\signflip_{0}$, which flips the ancilla qubit's state if and only if the path register is in the $\ket{\bm{0}}^{1}$ state.
Third, the threshold-dependent phase oracle within the Amplitude Amplification operator flips the ancilla state, if and only if the third register enables a profit larger than the current threshold.
That is, for known thresholds $T$ and unknown $p$, we have to implement the comparisons ``$p > T$''.

We realise all the above operations as multi-controlled single-target gates.
In \autoref{figure:ParallelisedMultiControlDecomposition} we provide an example of the optimal decomposition of a multi-controlled unitary $U$.
The control circuit is derived from the classical digital comparator \cite{Ercegovac2004DigitalArithmetic}:
Given two binary represented integers $a = a_{\ell} \ldots a_{1}$ and $b = b_{\ell} \ldots b_{1}$, introduce the Boolean variables
\begin{linenomath}
\begin{align}\label{equation:BooleanHelper}
    c_{i} \coloneqq a_{i} b_{i} \lor \overline{a}_{i} \overline{b}_{i},
\end{align}
\end{linenomath}
which are 1 if and only if $a_{i} = b_{i}$, respectively.
Then equality and strict inequalities in both directions can be expressed as
\begin{linenomath}
\begin{align}
    (a = b) &= \bigwedge_{i = 1}^{\ell} c_{i}, \label{equation:Equality} \\
    (a > b) &= \bigvee_{i = 1}^{\ell} \bigg(a_{i} \overline{b}_{i} \bigwedge_{j = i + 1}^{\ell} c_{j}\bigg), \label{equation:Greater} \\
    (a < b) &= \bigvee_{i = 1}^{\ell} \bigg(\overline{a}_{i} b_{i} \bigwedge_{j = i + 1}^{\ell} c_{j}\bigg). \label{equation:Lesser}
\end{align}
\end{linenomath}
If the integer $b$ is known, the variables \eqref{equation:BooleanHelper} can be reduced to a single literal, leaving \eqref{equation:Equality}--\eqref{equation:Lesser} in orthogonal disjunctive normal form (ODNF).
In addition, for $(a > b)$ only those clauses contribute for which $\overline{b}_{i}$ is 1, i.e., $b_{i}$ is 0.
Analogously, for $(a < b)$ only those clauses contribute for which $b_{i}$ is 1.

There are now two possibilities when implementing an operation $U$ which is controlled by a Boolean expression.
\begin{enumerate}
    \item $U$ is applied if and only if the Boolean expression is fulfilled.
    \item $U$ is first applied unconditionally.
    Then its inverse is applied if and only if the Boolean expression is not fulfilled.
\end{enumerate}
In any case, an ODNF that is checked for is beneficial as its quantisation is straightforward:
Each clause $\mathcal{C}$ translates into a multi-controlled-$U$ (resp.\ $U^{-1}$) where the control runs over all the qubits representing Boolean variables in $\mathcal{C}$.
If the classical literal is positive, we control on whether the corresponding qubit is in the $\ket{1}$ state, otherwise whether it is in the $\ket{0}$ state.
Because we assume an ODNF, we can ensure that $U$ (resp.\ $U^{-1}$) is applied at most once when translating all the clauses into consecutive multi-controlled operations.

\paragraph*{Gate count.}

For a given item weight $w_{m}$, we can now calculate the gate count for implementing the multi-controlled Hadamard gate $C_{\geq w_{m}}^{2}(\text{H}^{x_{m}}_{b})$.
Following the first strategy of applying H$^{x_{m}}_{b}$ if and only if the remaining capacity covers $w_{m}$ (that is if $\tilde{c} > w_{m} - 1$) and assuming \textbf{(i)} as well as \textbf{(iii)}, we obtain the count
\begin{linenomath}
\begin{align}\label{equation:FirstStrategyCapacityComparisonGates}
    \gatec_{\geq w_{m}}^{(1)} = \sum_{i = 1}^{\numbits{w_{m}}} (1 - (w_{m} - 1)_{i}) (2 (\numbits{c} - i) + 1).
\end{align}
\end{linenomath}
This implementation is favourable for weights $w_{m}$ with comparatively many ones in the less significant bits of the binary representation of the number $w_{m} - 1$.
Alternatively, following the second strategy, H$^{x_{m}}_{b}$ is first applied unconditionally (adding one to the gate count);
then its inverse is applied if and only if the remaining capacity is strictly smaller than $w_{m}$.
Again assuming \textbf{(i)} and \textbf{(iii)}, this yields a total gate count of
\begin{linenomath}
\HL{\begin{align}\label{equation:SecondStrategyCapacityComparisonGates}
    \gatec_{\geq w_{m}}^{(2)} = 1 + \sum_{i = 1}^{\numbits{w_{m}}} (w_{m})_{i} (2 (\numbits{c} - i) + 1).
\end{align}}
\end{linenomath}
For the concrete circuit construction, we can then simply pick the implementation that requires fewer gates, i.e.,
\begin{linenomath}
\HL{\begin{align}\label{equation:CapacityComparisonGates}
    \gatec_{\geq w_{m}} = \min\left(\gatec_{\geq w_{m}}^{(1)}, \gatec_{\geq w_{m}}^{(2)}\right).
\end{align}}
\end{linenomath}

The implementation of the reflection $\signflip_{0}$ is even more straightforward:
We simply have to check whether the path register is in the state $\ket{\bm{0}}^{1}$ and control the ancilla flip on the outcome.
According to \textbf{(i)} and \textbf{(iii)}, this yields a gate count of
\begin{linenomath}
\HL{\begin{align}\label{equation:ZeroComparisonGates}
    \gatec_{= 0} = 2 n - 1.
\end{align}}
\end{linenomath}

Lastly, the phase oracle is also realised as a single-qubit operation controlled on the truth value of a Boolean expression.
The derivation of the gate count follows the same construction as for the multi-controlled Hadamard gate, with the only exception being that the phase oracle asks for a strict inequality ``$p > T$''.
Hence we obtain
\begin{linenomath}
\HL{\begin{align}
    \gatec_{> T}^{(1)} &= \sum_{i = 1}^{\numbits{T}} (1 - (T)_{i}) (2 (\numbits{P} - i) + 1), \label{equation:FirstStrategyProfitComparisonGates} \\
    \gatec_{> T}^{(2)} &= 1 + \sum_{i = 1}^{\numbits{T}} (T + 1)_{i} (2 (\numbits{P} - i) + 1), \label{equation:SecondStrategyProfitComparisonGates} \\
    \gatec_{> T} &= \min\left(\gatec_{> T}^{(1)}, \gatec_{> T}^{(2)}\right). \label{equation:ProfitComparisonGates}
\end{align}}
\end{linenomath}

\paragraph*{Cycle count.}
\HL{Via the parallelised decomposition of multi-controlled gates exemplified in \autoref{figure:ParallelisedMultiControlDecomposition}, the respective cycle counts improve logarithmically over the gate counts.
More precisely, we obtain}
\begin{linenomath}
\HL{\begin{align}\label{equation:FirstStrategyCapacityComparisonCycles}
    &\cyclec_{\geq w_{m}}^{(1)} = \ \, \smashoperator{\sum_{i = 1}^{\numbits{w_{m}}}}\ (1 - (w_{m} - 1)_{i}) (2 \lceil\log_{2} (\numbits{c} - i)\rceil + 1)
\end{align}}
\end{linenomath}
\HL{for the first comparison strategy,}
\begin{linenomath}
\HL{\begin{align}\label{equation:SecondStrategyCapacityComparisonCycles}
    &\cyclec_{\geq w_{m}}^{(2)} = 1 + \ \, \smashoperator{\sum_{i = 1}^{\numbits{w_{m}}}}\ (w_{m} - 1)_{i} (2 \lceil\log_{2} (\numbits{c} - i)\rceil + 1)
\end{align}}
\end{linenomath}
\HL{for the second comparison method, and thus lastly}
\begin{linenomath}
\HL{\begin{align}\label{equation:CapacityComparisonCycles}
    \cyclec_{\geq w_{m}} = \min\left(\cyclec_{\geq w_{m}}^{(1)}, \cyclec_{\geq w_{m}}^{(2)}\right).
\end{align}}
\end{linenomath}
\HL{Meanwhile, the cycle cost for the reflection circuit $\signflip_{0}$ computes to}
\begin{linenomath}
\HL{\begin{align}\label{equation:ZeroComparisonCycles}
    \cyclec_{= 0} = 2 \lceil\log_{2}(n - 1)\rceil + 1.
\end{align}}
\end{linenomath}
\HL{For the phase oracle, we also list both implementation strategies together with their corresponding cycle count:}
\begin{linenomath}
\HL{\begin{align}
    \cyclec_{> T}^{(1)} &= \ \smashoperator{\sum_{i = 1}^{\numbits{T}}} (1 - (T)_{i}) (2 (\lceil\log_{2}(\numbits{P} - i)\rceil + 1), \label{equation:FirstStrategyProfitComparisonCycles} \\
    \cyclec_{> T}^{(2)} &= 1 + \smashoperator{\sum_{i = 1}^{\numbits{T}}} (T + 1)_{i} (2 \lceil\log_{2}(\lceil \numbits{P} - i)\rceil + 1), \label{equation:SecondStrategyProfitComparisonCycles} \\
    \cyclec_{> T} &= \min\left(\cyclec_{> T}^{(1)}, \cyclec_{> T}^{(2)}\right). \label{equation:ProfitComparisonCycles}
\end{align}}
\end{linenomath}

\subsubsection{\label{subsubsection:QTG}QTG}

After having detailed the resource estimations for all crucial subcircuits, we discuss now the overall gate and cycle count for the entire QTG.
We assume here the more resource-efficient version of the QTG that does not contain the last capacity update $U_{n}^{2}$.
\HL{Furthermore, because the ancilla qubits are used for controlling both the subtractions in the capacity register and the additions in the profit register, we may fuse the copy processes (and their inverses) that are used before applying the arithmetic rotation gates.
The gate and cycle count then is dominated by the operation that demands more rotations.}

\paragraph*{Gate count.}
The total gate count for QTG arises as the sum of the gates counts for the $n$ controlled Hadamard gates \HL{$U_{m}^{1}$ and for the capacity/profit updates with their copying steps merged as already mentioned.}
\begin{widetext}
\begin{linenomath}
\HL{\begin{align}\label{equation:QTGGates}
    \gatec_{\text{QTG}} &= \sum_{m = 1}^{n} \gatec_{\geq w_{m}} + 2 \gatec_{\text{QFT-PROFIT}} + 2 (n - 1) \gatec_{\text{QFT-CAPACITY}} \nonumber \\
    &\quad + 2\sum_{m = 1}^{n - 1} \big(\max(\numbits{P}, \numbits{c}) - \min(\text{LSO}(p_{m}), \text{LSO}(w_{m}))\big) + 2 (\numbits{P} - \text{LSO}(p_{n})) \nonumber \\
    &\quad + \sum_{m = 1}^{n - 1} \big(\numbits{P} - \text{LSO}(p_{m}) + \numbits{c} - \text{LSO}(w_{m}) + 2\big) + \numbits{P} - \text{LSO}(p_{n}) + 1.
\end{align}}
\end{linenomath}
\end{widetext}

\paragraph*{Cycle count.}
Up to this point, we have explored various methodologies to optimise the efficiency of gate and cycle counts for each component of the QTG. 
Now our focus shifts towards improving its overall cycle count.
We take into account the following considerations:
\begin{itemize}
    \HL{\item[I.] Copying qubit information from the path register into the ancilla register as well as the reversed operation can both be run in parallel with all the QFT$_{\text{c}}$ and QFT$^{\dagger}_{\text{c}}$ on the capacity register.
    \item[II.] The number of cycles within the QFT$_{\text{c}}$ (logarithmic in $c$) always exceeds the number of cycles involved in copying the control qubit onto the ancilla qubits (doubly-logarithmic in $c$ and $P$).}
    \item[III.] The QFT$_{\text{p}}$ on the profit register can be run in parallel with all the gates in the capacity register, as it does not involve any controlled operations.
\end{itemize}

Our first observation is, however, that the different layers $U^{\vphantom{1}}_{m} = U_{m}^{3} U_{m}^{2} U_{m}^{1}$ have to be executed mutually disjointly as the subtraction in $U_{m}^{2}$ and the addition in $U_{m}^{3}$ are controlled by the qubits that the biased Hadamard gates in $U_{m}^{1}$ act on.
Therefore, no components commute across different layers, and we obtain
\begin{linenomath}
\begin{align}\label{equation:QTGLayerwiseCycles}
    \cyclec_{\text{QTG}} = \sum_{m = 1}^{n} \cyclec_{m},
\end{align}
\end{linenomath}
where $\cyclec_{m}$ denotes the cycle count per layer unitary $U_{m}$.
\HL{Furthermore, there cannot be any parallelisation between $U_{m}^{1}$ and the rest of the layer;
though other parallelisation techniques are definitely applicable.}

\HL{In the first layer we parallelise, according to III, QFT$_{p}$ and QFT$_{c}$ as much as possible.
Due to I and II, we can also fit in the subcircuit copying the control qubit's information on the ancilla register.
If $\numbits{P} > \numbits{c}$, meaning that the $\cyclec_{\text{QFT-PROFIT}} > \cyclec_{\text{QFT-CAPACITY}}$, we can absorb the subtraction cycle, again due to III, into the profit QFT.
The remaining addition cycle as well as uncomputing the ancilla qubits can be absorbed into the subsequent QFT$_{\text{c}}^{\dagger}$.
In this first case we obtain}
\begin{linenomath}
\HL{\begin{align}\label{equation:QTGFirstBlockCycles1}
    \cyclec_{1}^{(1)} = \cyclec_{\geq w_{1}} + \cyclec_{\text{QFT-CAPACITY}} + \cyclec_{\text{QFT-PROFIT}}
\end{align}}
\end{linenomath}

\HL{Otherwise the subtraction cycle has to be carried out unparallelised, yielding}
\begin{linenomath}
\HL{\begin{align}\label{equation:QTGFirstBlockCycles2}
    \cyclec_{1}^{(2)} = \cyclec_{\geq w_{1}} + 2 \cyclec_{\text{QFT-CAPACITY}} + 1.
\end{align}}
\end{linenomath}

The other layer unitaries, except for the last one, are similarly constructed, but do not include any (inverse) QFT on the profit register.
In terms of cycles we are thus in the same situation as \eqref{equation:QTGFirstBlockCycles2}, that is
\begin{linenomath}
\HL{\begin{align}\label{equation:QTGSecondBlockCycles2}
    \cyclec_{m} = \cyclec_{\geq w_{m}} + 2 \cyclec_{\text{QFT-CAPACITY}} + 1.
\end{align}}
\end{linenomath}

The last layer unitary contains the inverse QFT for the profit register, but no weight subtraction in the capacity register.
Here, the copy subcircuit as well as the addition cycle cannot be executed in parallel with any other operation.
Therefore, we obtain
\begin{linenomath}
\HL{\begin{align}\label{equation:QTGLastBlockCycles}
    \cyclec_{n} &= \cyclec_{\geq w_{n}} + \lceil\log_{2}(\numbits{P} - \text{LSO}(p_{n}))\rceil \nonumber \\
    &\quad + \cyclec_{\text{QFT-PROFIT}} + 1.
\end{align}}
\end{linenomath}

\subsubsection{\label{subsubsection:QSearch}QMaxSearch}

In contrast to the QTG whose gate and cycle count can be determined in advance the \textbf{QMaxSearch} routine has to be simulated to infer its total resource requirement.
In order to derive a symbolic expression we assume that a run of \textbf{QMaxSearch} produces a sequence $(T_{k})_{k = 1}^{R}$ of intermediate thresholds with $T_{1}$ being the result of Integer Greedy.
The search method is therefore not able to find a better total profit than $T_{R}$ which might indicate that an optimal solution has been found.
For each of these thresholds $T_{k}$, \textbf{QSearch} is executed, incorporating the comparison ``$P_{\bm{x}} > T_{k}$'' as phase oracle.
Within a given instance of \textbf{QSearch} a sequence $(j^{(k)}_{\ell})_{\ell = 1}^{S(k)}$ of powers is constructed such that the termination condition
\begin{linenomath}
\begin{align}\label{equation:TerminationCriterion}
    \sum_{\ell = 1}^{S(R) - 1} 2 j^{(R)}_{\ell} + 1 < M \leq \sum_{\ell = 1}^{S(R)} 2 j^{(R)}_{\ell} + 1
\end{align}
\end{linenomath}
holds in the last application of \textbf{QSearch}.
For all prior iterations the termination condition is never triggered, because an improved threshold is found, respectively.

\paragraph*{Gate count.}
From the knowledge about the threshold $T_{k}$ and the corresponding power sequence we can now determine the gate count of the $k$-th iteration of \textbf{QSearch}:
Before applying $\QSop$ with a given power $j$ the QTG is applied once to create the initial superposition of feasible states.
The definition \eqref{equation:AmplitudeAmplificationOperator} of the Amplitude Amplification operator immediately gives rise to its gate count
\begin{linenomath}
\begin{align}\label{equation:AmplitudeAmplificationOperatorGates}
    \gatec_{\QSop_{k}} = 2 \gatec_{\text{QTG}} + \gatec_{= 0} + \gatec_{> T_{k}}.
\end{align}
\end{linenomath}
Multiplying the latter with the respective power and summing over all powers we obtain for the $k$-th iteration of \textbf{QSearch} a gate count of
\begin{linenomath}
\begin{align}\label{equation:QSearchGates}
    \gatec_{\text{QSearch}_{k}} = \sum_{\ell = 1}^{S(k)} (2 j_{\ell}^{(k)} + 1) \gatec_{\text{QTG}} + j_{\ell}^{(k)} (\gatec_{= 0} + \gatec_{> T_{k}}).
\end{align}
\end{linenomath}
The total gate count of \textbf{QMaxSearch} is then given by
\begin{linenomath}
\begin{align}\label{equation:QMaxSearchGates}
    \gatec_{\text{QMaxSearch}} = \sum_{k = 1}^{R} \gatec_{\text{QSearch}_{k}}.
\end{align}
\end{linenomath}

\paragraph*{Cycle count.}
Because the operators $\signflip_{0}$ and $\signflip_{P_{\bm{x}} > T_{k}}$ are controlled on the profit register's qubits no further parallelisation within $\QSop$ is possible.
Therefore, we obtain for the \textbf{QSearch} cycle count
\begin{linenomath}
\begin{align}\label{equation:QSearchCycles}
    \cyclec_{\text{QSearch}_{k}} = \sum_{\ell = 1}^{S(k)} (2 j_{\ell}^{(k)} + 1) \cyclec_{\text{QTG}} + j_{\ell}^{(k)} (\cyclec_{= 0} + \cyclec_{> T_{k}})
\end{align}
\end{linenomath}
and, analogously, for the final cycle count of the entire 
\textbf{QMaxSearch} routine
\begin{linenomath}
\begin{align}\label{equation:QMaxSearchCycles}
    \cyclec_{\text{QMaxSearch}} = \sum_{k = 1}^{R} \cyclec_{\text{QSearch}_{k}}.
\end{align}
\end{linenomath}
The formula \eqref{equation:QMaxSearchCycles} is precisely what is being used for calculated the elapsed quantum cycles in \autoref{figure:QTGRuntimeBenchmark}.

\subsection{\HL{Benchmarking of competitive quantum methods}}\label{Section:BenchmarkingOfCompetitiveMethods}

\HL{This section contains a brief overview and runtime estimations of the nested quantum search \cite{Cerf2000NestedQuantumSearchAndStructuredProbleems} and quantum branch-and-bound \cite{Montanaro2020QuantumSpeedupOfBranchAndBoundAlgorithms, Chakrabarti2022UniversalQuantumSpeedupForBranchAndBoundBranchAndCutAndTreeSearchAlgorithms} algorithms, together with our benchmarking reasoning.}

\subsubsection{\HL{Nested Quantum Search}}

\HL{Nested quantum search (\textbf{NQS}) is a nested version of Grover's algorithm specifically tailored to constraint satisfaction problems, where the search Hilbert space is decomposed into two factors $\hil_{\text{S}} = \hil_{\text{A}} \otimes \hil_{\text{B}}$, corresponding to a decomposition $\text{S} = \text{A} \cup \text{B}$ of the set of classical variables. 
The algorithm itself is a variant of \textbf{QSearch}, where the superposition-creating unitary $\mathcal{A}$ consists of the following two steps:
\begin{enumerate}
    \item Grover's search in the first tensor factor $\hil_{\text{A}}$ to (nearly) obtain a uniform superposition of \emph{could-be} solutions, i.e. partial assignments (only variables in A are assigned) that fulfill all constraints that are testable on these variables.
    Accordingly, these are precisely the partial assignments that \emph{could be} extended to a full solution of the problem.
    The employed oracle checks whether a partial assignment is a could-be solution and is assumed to only act on $\hil_{\text{A}}$.
    \item Subsequent Grover's search on the second tensor factor $\hil_{\text{B}}$, but with an oracle that acts on the entire space $\hil_{\text{S}}$ and marks full (and not partial) solutions.
\end{enumerate}}

\HL{The first step of \textbf{NQS} creates a superposition of all, say $t_{\text{A}}$, could-be solutions plus a vanishing amount of non could-be solutions;
this step involves $\sim \sqrt{2^{\abs{\text{A}}} / t_{\text{A}}}$ Grover iterations.
Second, assume that each could-be solution gives rise to at most one full solution and let $t_{\text{B}} \in [0, 1]$ be the average number of full solutions per partial solution.
Then the total number of solutions is given by $t_{\text{AB}} = t_{\text{A}} t_{\text{B}} \leq t_{\text{A}}$.
Executing the second step produces a superposition of all solutions with amplitude $\sim 1 \sqrt{t_{\text{A}}}$ and all infeasible solutions, which extend from could-be solutions that did not give rise to any solution, with amplitude $\sim 1 / \sqrt{t_{\text{A}} 2^{\abs{\text{B}}}}$, plus a vanishing amount of other infeasible solutions after $\sim \sqrt{2^{\abs{\text{B}}} / t_{\text{B}}}$ steps.
The total number of Grover iterations per state preparation is}
\begin{linenomath}
\HL{\begin{align*}
    T \sim \sqrt{2^{\abs{\text{A}}} / t_{A}} + \sqrt{2^{\abs{\text{B}}} / t_{B}}.
\end{align*}}
\end{linenomath}

\HL{Because there are now $t_{\text{AB}}$ solution states each having an amplitude of $\sim 1 / \sqrt{t_{\text{A}}}$, they can be amplified to yield a positive outcome upon measuring after $\sim \sqrt{t_{\text{A}} / t_{\text{AB}}}$ iterations within \textbf{QSearch}.
In total this yields}
\begin{linenomath}
\HL{\begin{align}\label{equqation:NQSRunTIme}
    T_{\text{\textbf{NQS}}} = T \sqrt{t_{\text{A}} / t_{\text{AB}}} \sim \sqrt{2^{\abs{\text{A}}} / (t_{\text{A}} t_{\text{B}})} + \sqrt{2^{\abs{\text{B}}}}
\end{align}}
\end{linenomath}
\HL{Grover iterations.}

\HL{\textbf{NQS} can also be used for constrained optimisation problems by employing it as a feasible state preparation subroutine (i.e. the measurement is omitted), on top of which \textbf{QMaxSearch} is executed.
For the $\KP$, we can utilise the constraint checking techniques from the QTG also within \textbf{NQS}.
To derive the trend of Grover iterations required to find an optimum, we first find the best decomposition of the $n$ decision variables into variable sets A and B:
Following \cite{Pisinger2005WhereAreTheHardKnapsackProblems}, we define a \textit{break item}}
\begin{linenomath}
\HL{\begin{align}
    h = \max\left\{ m': \sum_{m = 1}^{m'} w_{m'} \leq c \right\} ,
\end{align}}
\end{linenomath}
\HL{where every partial assignment up to this item is feasible.
The first set of variables will be $\text{A} = \{1, \ldots, h, \ldots, h + k\}$ with $0 \leq k \leq n - h\}$ yet to be determined;
accordingly, B will be given as $\text{B} = \{h + k + 1, \ldots, n\}$.
Furthermore, note that every feasible partial assignment can always be extended to a feasible full assignment and that any lower bound on the number of feasible solutions also gives a lower bound on the runtime of \textbf{QMaxSearch} as fewer elements have to be searched through for the optimum.
Thus, assuming that each partial assignment has exactly one feasible solution and therefore that $t_{\text{B}} = 1$ and $t_{\text{AB}} = t_{\text{A}}$ gives an optimistic lower bound on the runtime of \textbf{QMaxSearch}.}

\HL{Using our assumption and the structure of A and B, \eqref{equqation:NQSRunTIme} gives rise to an \textbf{NQS}-runtime of}
\begin{linenomath}
\HL{\begin{align*}
    T_{\text{\textbf{NQS}}} \sim \sqrt{2^{h + k} / 2^{h}} + \sqrt{2^{n - h - k}} = \sqrt{2^{k}} + \sqrt{2^{n - h - k}}
\end{align*}}
\end{linenomath}
\HL{This produces an equal superposition of $2^{h}$ feasible solutions upon which we now have to find the maximum which therefore requires $\sim \sqrt{2^{h}}$ iterations of \textbf{NQS} within \textbf{QMaxSearch}, yielding a total number of Grover iterations}
\begin{linenomath}
\HL{\begin{align*}
    T_{\text{\textbf{NQS}}} \sqrt{2^{h}} \sim \sqrt{2^{h + k}} + \sqrt{2^{n - k}}.
\end{align*}}
\end{linenomath}
\HL{This expression is minimal for $k = \frac{n - h}{2}$, leading to a number of Grover iterations of order}
\begin{linenomath}
\HL{\begin{align}\label{equation:OptimalNQSRuntime}
    \sqrt{2^{(n + h) / 2}} + \sqrt{2^{n - h - (n - h) / 2}} = 2^{(n + h) / 4}.
\end{align}}
\end{linenomath}

\begin{figure}
    \centering
    \includegraphics[width=0.48\textwidth]{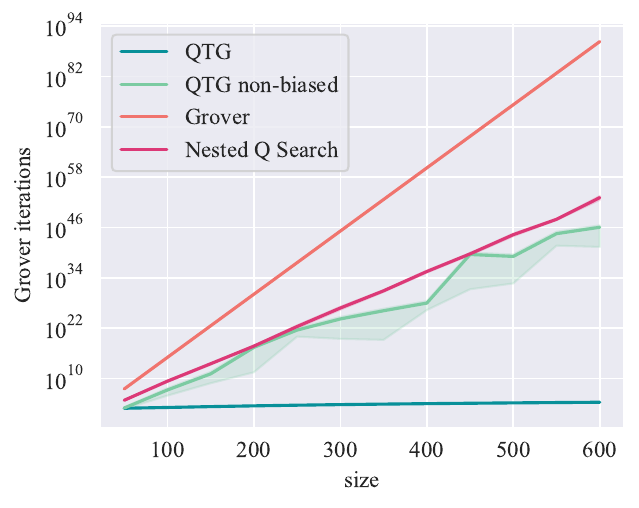}
    \caption{\HL{Comparison between the number of Grover iterations required by QTG-based search with and without bias, Grover's algorithm, and \textbf{NQS}.
    We observe that the number of Grover iterations is almost constant for the biased QTG-based search.
    For the other methods, we have a linear increase with the problem size: for Grover's search for problems of size 400 the algorithm already requires more than $10^{50}$ oracle calls;
    non-biased QTG-based search and \textbf{NQS} perform better, but still requiring more than $10^{20}$ oracle calls.
    Interestingly, the optimistic lower bound on the runtime of \textbf{NQS} seems to match the worst-case behaviour of the non-biased QTG-based search.}}
    \label{figure:GroverIterationsComparison}
\end{figure}

\HL{It is important to notice that setting $t_{\text{B}} = 1$ is a very strong assumption (it makes the 0-1-$\KP$ instance trivial), and we use it only to gain insights into the performance of \textbf{NQS}.
The benchmark results on real instances are presented in \autoref{figure:GroverIterationsComparison}.}

\subsubsection{\HL{Quantum branch-and-bound}}\label{Section:Qbnb}

\HL{Quantum branch-and-bound (QBnB) \cite{Montanaro2020QuantumSpeedupOfBranchAndBoundAlgorithms, Chakrabarti2022UniversalQuantumSpeedupForBranchAndBoundBranchAndCutAndTreeSearchAlgorithms} is one of the most promising candidates for solving ILPs, having a provable worst-case asymptotic near-quadratic speed\-up over its classical counterpart.
The near-quadratic speed\-up is achieved by using quantum walks \cite{Apers2019AUnifiedFrameworkOfQuantumWalkSearch} on the tree explored by the classical algorithm. 
The method has two main ingredients:
\begin{enumerate}
    \item Quantum tree size estimation \cite{Ambainis2017QuantumAlgorithmForTreeSizeEstimation}
    to get an upper bound on the size of the tree. This implementation shows how to get a quadratic improvement over a classical backtracking algorithm which evaluates $T$ nodes of a search tree.
    \item Quantum tree search \cite{Apers2019AUnifiedFrameworkOfQuantumWalkSearch} to return the label of the marked node if the tree contains at least one marked node, or ``not found'' if there is no marked node.
\end{enumerate}}

\HL{In general, the QBnB routine requires multiple applications of quantum tree size estimation and quantum tree search.
At the core of these subroutines lies quantum backtracking \cite{Montanaro2018QuantumWalkSpeedupOfBacktrackingAlgorithms, Martiel2020PracticalImplementationQuantumBacktrackingAlgorithm, Ambainis2017QuantumAlgorithmForTreeSizeEstimation}.
In the following, we give an optimistic runtime estimation of the QBnB method.
We benchmark the QBnB for the 0-1-$\KP$ under the following assumptions: 
\begin{enumerate}
    \item The optimal solution and the respective tree size $T$ are known, hence quantum tree size estimation has to be applied only once.
    \item The success probability of the quantum subroutines is 1.
    \item The most expensive part of the algorithm is the backtracking, hence we consider all the other quantum subroutines to be queried once.
\end{enumerate}}

\HL{In general, in order to find the optimal solution, the expected number of oracles calls is at least $\sqrt{T n}$, where $T$ represents the number of nodes that are actually explored by the corresponding classical BnB algorithm, and $n$ is the number of items.
To justify this value, we give a summary of the backtracking algorithm together with our reasoning. 
Consider a rooted tree with $T$ vertices $r, 1, \ldots, T - 1$, where $r$ is the root. 
$A$ denotes the set of vertices at an even distance from $r$, and $B$ the set of vertices at an odd distance from $r$. $x \rightarrow y$ signifies that $y$ is a child of $x$ in the tree.
The quantum walk operates in the space spanned by $\{\ket{x} \defcolon x \in \{r, 1, \ldots, T - 1\}\}$ and starts in the state $\ket{r}$. 
This quantum walk utilises a set of diffusion operators $D_{x}$, where $D_{x}$ is the identity operator if $x$ is a solution; otherwise, it diffuses on the subspace spanned by $\{\ket{x}\} \cup \{\ket{y} \defcolon x \rightarrow y\}$. 
A step of the quantum walk involves applying the operator $R_{B} R_{A}$, where}
\begin{linenomath}
\HL{\begin{align*}
    R_{A} = \bigoplus_{x \scriptin A} D_{x} \text{ and } \ketbra{r}{r} + R_{B} = \bigoplus_{x \scriptin B} D_{x}.
\end{align*}}
\end{linenomath}
\HL{Using these operators, a marked vertex is detected, because the phase estimation operator $R_{B} R_{A}$ enables the traversing of paths leading to a solution in the tree.
According to \cite{Montanaro2018QuantumWalkSpeedupOfBacktrackingAlgorithms} the operator $R_{B} R_{A}$ is applied to precision $\beta / \sqrt{T n}$.}

\HL{For our benchmarking results, we assume that $\beta = 1$, implying that the application of the phase estimation operator has to be applied exactly $\sqrt{T n}$ times.
We execute the classical algorithm and record the tree size $T$.}

\HL{Within the BnB routine, we use the commonly used upper bound heuristic \emph{Fractional Greedy} \cite{Kellerer2004KnapsackProblems} which, similar to (Integer) Greedy, sorts all items by decreasing density and then greedily packs items into the knapsack until an item does not fit into the knapsack (by definition the successor of the break item $h$).
It then returns the sum of all the profits of contained items plus the profit of the fraction of the item that could be included additionally, i.e.}
\begin{linenomath}
\HL{\begin{align*}
    \sum_{m = 1}^{h} p_{m} + p_{h + 1} \frac{c - \sum_{m = 1}^{h} w_{m}}{w_{h + 1}}.
\end{align*}}
\end{linenomath}

\HL{Because this routine is very similar to the (Integer) Greedy method, all the previous calculations for the cycle count of the QTG circuit can be used to calculate the costliness of the quantum implementation of Fractional Greedy.}

\end{document}